\documentclass[%
 reprint,
superscriptaddress,
 amsmath,amssymb,
 aps,
]{revtex4-2}
\usepackage{siunitx}
\usepackage{booktabs} 
\usepackage{amsmath}
\usepackage{array}
\usepackage{graphicx} 
\usepackage{epstopdf}
\usepackage{hyperref}
\usepackage{siunitx}
\hypersetup{colorlinks=true}
\newcommand\Rm{{\text{Rm}}}
\newcommand\Ro{{\text{Ro}}}
\usepackage{caption}
\usepackage{subcaption}

\usepackage{xcolor}
\usepackage{dcolumn}
\newcommand{\tbc}[1]{\textcolor{red}{#1}}
\usepackage{bm}
\usepackage{booktabs}
\usepackage{dashrule}

\newcommand\Ra{{\text{Ra}}}
\newcommand\E{{\text{E}}}
\newcommand\Pm{{\text{Pm}}}
\newcommand\Prr{{\text{Pr}}}
\renewcommand{\Re}{\text{Re}}
\newcommand{\Ebax}{\ensuremath{E_\text{b}^\text{ax}}}
\newcommand{\Ebdip}{\ensuremath{E_\text{b}^\text{dip}}}
\newcommand{\Eb}{\ensuremath{E_\text{b}}}
\newcommand{\Ek}{\ensuremath{E_\text{k}}}
\newcommand{\blackdiamondshape}{{\rotatebox[origin=c]{45}{$\blacksquare$}}}
\definecolor{pygreen}{HTML}{2ca02c}
\usepackage{relsize}

\begin{document}

\preprint{APS/123-QED}

\title{Large $\Pm$ small-scale kinematic dynamo in protoneutron stars}

\author{Shipra Verma}
\email{shipra97verma@gmail.com}
\affiliation{Department of Physics, Indian Institute of Technology Kharagpur, West Bengal, 721302, India}
\author{Kannabiran Seshasayanan}
\affiliation{Department of Applied Mechanics $\&$ Biomedical Engineering, Indian Institute of Technology Madras, Chennai, 600036, India}
\author{Raphaël Raynaud}
\affiliation{Université Paris Cité, Université Paris-Saclay, CEA, CNRS, AIM, F-91191 Gif-sur-Yvette, France}
\author{Jérôme Guilet}
\affiliation{Université Paris-Saclay, Université Paris Cité, CEA, CNRS, AIM, F-91191 Gif-sur-Yvette, France}

\begin{abstract}
Magnetars are young, isolated neutron stars that possess an exceptionally strong magnetic field, with surface dipolar strengths on the order of $\SI{e15}{G}$. One of the plausible scenarios for generating such a strong field is an exponential amplification by a turbulent convective dynamo during the protoneutron star phase. However, the short expected duration of the convection ($\sim \SI{10}{s}$) imposes a stringent constraint on the dynamo growth rate. We perform an extensive set of 82 three-dimensional convective dynamo simulations in the anelastic approximation and investigate the kinematic phase to quantify the dynamo growth rate $\gamma$. We find that $\gamma$ increases with both the magnetic Prandtl number $\Pm$ and the Rayleigh number $\Ra$, with the most unstable mode becoming highly non-axisymmetric and multipolar. We further observe a gradual transition from large-scale to small-scale dynamo as the magnetic Reynolds number $\Rm$ increases, resulting in a magnetic field that is predominantly concentrated at small scales. The trend remains unchanged when the outer magnetic boundary condition is varied. Since resolving the increasingly small scales becomes numerically impractical, we employ the theoretical small-scale Kazantsev dynamo model to explore the large $\Pm$ regime characteristic of protoneutron stars. The model qualitatively captures the growth rate behaviour observed in simulations and, upon extrapolation to the large $\Pm$ limit, indicates that $\gamma$ is only weakly dependent on the resistivity in a PNS. Under conditions relevant to the PNS, this model predicts a magnetic energy growth rate of the order of $\sim \SI{1}{\per\milli\second}$.
\end{abstract}

\maketitle

\section{Introduction}

When a massive main-sequence star with an initial mass $M \gtrsim \SI{8}{M_\odot}$ \cite{ugliano2012progenitor,muller2025minimum,maraston2025stellar} undergoes core-collapse supernova, it leaves behind a compact object, either a black hole or a neutron star. In the latter case, the remnant initially forms a protoneutron star (PNS). The PNS consists of two primary regions: the rebounded inner core and the surrounding mantle. The mantle is neither in thermal nor in $\beta$-equilibrium and therefore undergoes rapid cooling and deleptonization. Convection in this region, driven by a negative radial gradient of entropy and lepton fraction \cite{Epstein1979convection}, enhances the heat and lepton transport \cite{Keil1996LedouxConvectionPNS, Nagakura2019PNSConvection}. Gradually, equilibrium is established, ultimately leading to the formation of a cold, stable neutron star (NS) \cite{pascal2021modelling,cerda2019neutron}.

Among the diverse types of NSs observed to date \cite{harding2013neutron}, magnetars represent a particularly distinct class. They are young, slowly rotating, and possess exceptionally strong magnetic fields, with surface dipolar strengths on the order of $\SI{e15}{G}$ \cite{Olausen2014McGillMagnetarCatalog, kaspi2017magnetars,neutronReiss}. The formation of a rapidly rotating magnetar can also power some of the most energetic stellar transients, including hypernovae, long gamma-ray bursts and super-luminous supernovae \cite{Duncan1992Magnetar, Woosley2010MagnetarGRB, Kasen2010MagnetarSLSN, Metzger2011MagnetarModel}. However, the origin of their intense magnetic fields remains an important open question, facing the stringent observational constraint that $10-40\%$ of newborn neutron stars are magnetars \cite{kouveliotou1999magnetars, beniamini2019formation}. One scenario is based on the simple assumption of magnetic flux conservation \cite{schneider2019stellar,ferrario2006modelling}, which requires the iron core progenitor to have a very strong magnetic field of the order of $10^{10}-\SI{e11}{G}$. It is highly uncertain whether a large enough number of progenitors can have such strong magnetic fields \cite{makarenko2021testing}. Moreover, it is unclear if the fossil field is left undisturbed when subjected to the strong turbulent convective motions in PNS.

A second class of scenarios for magnetar formation relies on the amplification of the magnetic field by dynamo action during the first instants following the PNS formation \cite{spruit2008origin}. Several dynamo mechanisms \cite{dehman2025magnetar} have now been proposed to develop in the PNS stage, notably the convective dynamo \cite{Thompson1993NSDynamos, raynaud2020magnetar, white2022origin, masada2022convection}, the magnetorotational instability \cite{obergaulinger2009semi, mosta2015large, guilet15, reboul2021global}, and the Tayler-Spruit dynamo triggered by fallback \cite{barrere2022new,barrere2023,barrere2025tayler, igoshev2025connection}. Collectively, these studies suggest that a strong, large-scale magnetic field consistent with the observed surface dipolar field in magnetars can be generated if the rotation rate is sufficiently fast. However, the actual efficiency of the dynamo mechanisms under conditions characteristic of a PNS remains highly uncertain, because numerical models cannot capture the full range of length-scales spanned by magnetohydrodynamic turbulence. In particular, PNSs are characterised by a very large value of the magnetic Prandtl number (the ratio of kinematic viscosity to magnetic diffusivity, $\Pm\sim10^{13}$), whose description is especially challenging due to the scale separation between the velocity and the resistive scales \cite{guilet2022mri,held2022}. Lander \cite{lander2021generating} has argued that the regime of large Pm leads to very slow dynamo action due to inefficient reconnection rates, thereby questioning the efficiency of the aforementioned large-scale dynamos. Additionally, strong convective motions are expected to be sustained for only $\sim \SI{10}{s}$ \cite{Roberts2012PRL} after the PNS formation, placing a stringent upper limit on the duration of convective dynamo. 

To clarify the growth timescale of the convective dynamo and its dependence on diffusive processes, we perform an extensive set of 3D global convective dynamo simulations in the anelastic approximation. We focus on the kinematic phase and determine the variation of characteristic growth properties with the control parameters. Additionally, we employ a theoretical framework to extrapolate our findings to the large $\Pm$ regime. A detailed description of the physical model, numerical setup, and methodology used is given in Section \ref{sec: model}. The results on the magnetic energy growth rate, along with visualisation and quantification of the magnetic field structure, are discussed in Section \ref{sec: res}. A theoretical model and its extrapolation to the large $\Pm$ limit, aimed at quantifying the growth rate of magnetic energy in PNS, are presented in Section \ref{sec: KK}. Concluding remarks are given in Section~\ref{sec: conclusion}.

\section{PNS interior, Governing equations, and Methodology} \label{sec: model}

\subsection{Governing equations}
Following Raynaud et al. (2020) \cite{raynaud2020magnetar}, we model the convective zone of a $\SI{1.78}{M_\odot}$ PNS $\SI{0.2}{s}$ after the core bounce as an electrically conducting fluid subject to an initial seed magnetic field in a rotating spherical shell with an inner boundary at $r=r_i$ and an outer boundary at $r=r_o$. Convective motions are driven and maintained by a constant heat flux at the inner and outer shell boundaries. Assuming the background is steady and isentropic, such that the superadiabatic temperature gradient is very small, we can filter out the high-frequency sound waves. This is known as the anelastic approximation \cite{Lantz1999AnelasticMHD}, which enables us to model convection in PNS as a first-order perturbation to the background. The radial profiles for the background corresponding to density ${\rho(r)}$, temperature ${T(r)}$, gravity ${g(r)}$, magnetic diffusivity ${\eta(r)}$, thermal and viscous neutrino diffusivity ${\kappa(r)}$ and ${\nu(r)}$, respectively, are adopted from the 1D core-collapse supernova simulation by Hüdepohl \cite{dissertation}, as done in \cite{raynaud2020magnetar}. We non-dimensionalise the governing equations using the shell width $d =r_o - r_i$ as the characteristic length scale. The velocity $\vec{v}$ is scaled by ${\nu}_{r_o}/d$, the entropy $s$ by $\left( {\partial s}/{\partial r}\right)_{r_o} d$, the pressure $p$ using ${\rho}_{r_o} {\nu}_{r_o} \Omega$, and the magnetic field $\vec{B}$ with $({\rho}_{r_o} \mu_0 {\eta}_{r_o} \Omega)^{1/2}$, where $\mu_0$ is the magnetic permeability and $\Omega$ is the PNS rotation rate. Quantities with subscript $r_o$ denote their values evaluated at $r=r_o$. Within this framework, we solve the anelastic magneto-hydrodynamic dimensionless equations in the corotating frame, as given in Eqs. \eqref{eqn: fMHD}-\eqref{eqn: lMHD}, where, $Q^{\nu}$ $(=\sigma_{ij}\partial_j v_i)$ is heating due to viscous forces $\vec{F}^{\nu}$ $ (F^{\nu}_i= \tilde{\rho}^{-1}\partial_j \sigma_{ij})$ and $\sigma_{ij} = \tilde{\rho} \tilde{\nu}\left( \partial_i v_j + \partial_j v_i - 2\partial_k v_k \delta_{ij}/3\right)$ is the stress tensor with $\delta_{ij}$ as the Kronecker delta. Here, tilde denotes a background quantity normalised by its respective value at the outer shell of the convective zone.
\begin{widetext} 
\begin{gather}\label{eqn: fMHD}
\vec{\nabla}\cdot\left(\tilde{\rho}\vec{v}\right)=0,\\
\vec{\nabla}\cdot\vec{B}=0,\\
    \dfrac{\partial \vec{B}}{\partial t} = \vec{\nabla} \times \left( \vec{v}\times\vec{B}\right)-\dfrac{1}{\Pm}\vec{\nabla}\times\left(\tilde{\eta}\,\vec{\nabla}\times\vec{B}\right),\\
\tilde{\rho}\tilde{T}\left(\dfrac{\partial s}{\partial t} +
\vec{v}\cdot\vec{\nabla} s\right) =
\dfrac{1}{\Prr}\vec{\nabla}\cdot\left(\tilde{\kappa}\tilde{\rho}\tilde{T}\vec{\nabla} s\right) +
\dfrac{\Prr}{\Ra}Q_\nu
+\dfrac{\Prr}{\Pm^2\,\E\,\Ra}\tilde{\eta}\left(\vec{\nabla}
\times\vec{B}\right)^2,\\
\left(\dfrac{\partial \vec{v}}{\partial t}+\vec{v}\cdot\vec{\nabla}\vec{v}\right)
= -\frac{1}{\E}\vec{\nabla}\left({\dfrac{p}{\tilde{\rho}}}\right) - \dfrac{2}{\E}\vec{e}_z\times\vec{v}
- \dfrac{\Ra}{\Prr} \dfrac{d \tilde{T}}{d r} \,s\,\vec{e}_r
+\dfrac{1}{\Pm\,\E}\dfrac{1}{\tilde{\rho}}\left(\vec{\nabla}\times \vec{B}
\right)\times \vec{B}+ \vec{F}^{\nu} \label{eqn: lMHD}.
\end{gather}
\end{widetext}
The four dimensionless numbers which govern this set of equations are the Rayleigh ($\Ra$), Ekman ($\E$), magnetic Prandtl ($\Pm$), and thermal Prandtl ($\Prr$) numbers, respectively, defined as,
\begin{gather}
    \Ra=\dfrac{{T}_{r_o}d^3 \left(\frac{\partial s}{\partial r}\right)_{r_o}} {{\nu} _{r_o} {\kappa} _{r_o} }; \mbox{ } \E=\dfrac{{\nu} _{r_o}}{\Omega d^2};\\
    \Pm=\dfrac{{\nu} _{r_o}}{{\eta}_{r_o}}; \mbox{ } \Prr=\dfrac{{\nu} _{r_o}}{{\kappa} _{r_o}}.
\end{gather}
An important non-dimensional number in our study is the modified Rayleigh number
\begin{equation}
\Ra^*= {\Ra}\frac{\E^2}{\Prr}\,,
\end{equation}
a measure of the ratio of buoyancy to Coriolis force in the momentum equation. It should be noted that $\Ra^*$ is independent of neutrino diffusivities, $\nu$ and $\kappa$.

Other diagnostic quantities are the Reynolds and magnetic Reynolds numbers, respectively, defined as
\begin{equation}
    \Re = \frac{Ud}{\nu_{r_o}} \quad\text{and}\quad
    \Rm = \frac{Ud}{\eta_{r_o}} = \Re \Pm, 
\end{equation}
and the Rossby number 
\begin{equation}
    \Ro = \frac{U}{\Omega d} = \Re \E,
\end{equation}
where $U$ is the root mean square velocity measured in the simulations. Additionally, we introduce a modified Rossby number \cite{ChristensenAubert2006}
\begin{equation}
    \Ro_{\ell} = \Ro \, \dfrac{{\bar{\ell}_\text{v}}}{\pi}.
\end{equation}
Here, $\bar{\ell}_\text{v}$ is the mean characteristic angular wavenumber of the flow defined as
\begin{equation}
    \bar{\ell}_\text{v}= \dfrac{\sum \ell \mathcal{E}_{\ell}}{\sum \mathcal{E}_{\ell}},
\end{equation}
where, $\mathcal{E}_{\ell}$ is the time-averaged kinetic energy in the angular wavenumber $\ell$, and the summation $\sum$ is over $\ell \in [0,\mbox{ }\ell_{\text{max}}]$.

Following the conventional approach for modelling free surfaces, we use the stress-free boundary condition for the velocity field at $r_i$ and $r_o$. The heat flux is imposed through a fixed entropy gradient across boundaries. For most of our simulations, we take perfectly conducting magnetic boundary conditions (that is, vanishing radial magnetic field and tangential electric field) at the inner and outer boundaries owing to the very high electrical conductivity in PNS. Additional simulations were performed with pseudo-vacuum boundary conditions (that is, vanishing tangential magnetic field) to check the dependence of the results on the outer magnetic boundary conditions. To rescale the simulations to physical units, we follow \cite{raynaud2020magnetar} wherein $\rho_{r_o} = 8.3\times 10^{12} \mbox{ }\si{\gram\per\centi\cubic\meter}$, $\nu_{r_o} = 4.67\times 10^{12} \mbox{ }\si{\centi\square\meter\per\second}$, $\kappa_{r_o} = 4.67\times 10^{13} \mbox{ }\si{\centi\square\meter\per\second}$, $\eta_{r_o} = 2.33\times 10^{12} \mbox{ }\si{\centi\square\meter\per\second}$, $d=12.5 \mbox{ }\si{\kilo\meter}$, $r_o = 25 \mbox{ }\si{\kilo\meter}$ and the fixed heat flux at outer boundary $\Phi_{r_o} = 2 \times 10^{52} \mbox{ }\si{\text{erg}\per\second}$.

\subsection{Numerical methods}

The numerical solver used to solve Eqs. \eqref{eqn: fMHD}-\eqref{eqn: lMHD} is the benchmarked pseudo-spectral code \href{https://magic-sph.github.io/}{MagIC} \cite{gastine2012effects,schaeffer2013efficient}, which employs a poloidal and toroidal decomposition of the fields. The basis functions for expansion are Chebyshev polynomials in the radial direction and spherical harmonics in the angular direction. The time integration scheme used in the kinematic phase is the Semi-implicit S-DIRK of 3$^{\text{rd}}$ order. Simulations were performed using a grid resolution ranging from $(n_r,\mbox{ }n_{\theta},\mbox{ }n_{\phi})={ }(145,\mbox{ }160,\mbox{ }320)$ to $(n_r,\mbox{ }n_{\theta},\mbox{ }n_{\phi})=(257,\mbox{ }640,\mbox{ }1280)$. The resolution is increased as $\Pm$ and $\Ra$ are increased to ensure that the viscous and resistive scales are resolved. We performed convergence tests to check that the results are independent of resolution. Additionally, tests were conducted using different numerical schemes, such as finite differences in the radial direction and Semi-implicit S-DIRK of 2$^{\text{nd}}$ order in time, and the results remained unchanged. 

\subsection{Parameter study}

We perform a parameter study of 82 global convective dynamo simulations and investigate the kinematic dynamo problem. In this work, we analyse the growth rate of magnetic energy along with the topology of the generated field, and describe the driving mechanism of its generation. The focus is on the regime relevant to protoneutron stars, characterised by a rotation period of a few milliseconds, and $\Re \sim 10^{4}$ and $\Pm \sim 10^{13}$ \cite{lander2021generating}, corresponding to a relatively low neutrino viscosity and a much lower magnetic diffusivity. Due to numerical constraints, simulating the system at such high $\Re$ and large $\Pm$ is not feasible. Instead, we conduct studies at more accessible parameter values and then extrapolate to the regime of interest. 

Higher Reynolds numbers can be achieved by increasing $\Ra$ while keeping either $\E$ or $\Ra^*$ fixed. To examine both cases, we begin with a baseline simulation corresponding to a specific value of $\Ra$ and $\E$ (first row in red in Table \ref{tab: runs}). In the first case, $\E$ is fixed while $\Ra$ is increased. This fixes the ratio of viscous to Coriolis force, such that the buoyancy forcing is increased, implying an increase in $\Ro$. In the second case, $\Ra^*$ is fixed while $\Ra$ is increased and $\E$ is decreased simultaneously. This fixes the ratio of buoyancy to Coriolis forces, implying that $\Ro$ remains approximately constant, while buoyancy forcing increases compared to diffusive effects. In both cases, the behaviour with increasing $\Pm$ is studied, reaching moderately high $\Pm$ values. Additional simulations have been conducted to check the dependence of the results on the outer magnetic boundary condition. For all simulations, we take $\Pr = 0.1$, while $\Pm \in [2,35]$, $\E \in [0.73 \times 10^{-3},1.66\times10^{-3}]$, and $\Ra \in [5.8\times 10^3, 3.0\times 10^4]$ which translates to $\Re \in [42,101]$, $\Rm \in [8.4 \times 10^1,3.0 \times 10^3]$ and $\Ro \in [0.07, 0.15]$ -- see Table \ref{tab: runs}.

\begin{table*}
\centering
\caption{Index of Runs}
\vspace{-0.15in}
\caption*{\footnotesize For all simulations $\Pr =0.1$ and the shell aspect ratio $r_i/r_o = 0.5$. $\textbf{B}_{\text{bc}} = \{2,4\}$ represents a perfect conductor or pseudo-vacuum outer magnetic boundary condition, respectively, and the corresponding simulations are separated using a solid line. $\Pm=2-35$ denotes a range of values of $\Pm \mbox{ }(= 2,3,4,6,8,10,15,20,25,30,35)$, and $P$ is the rotation period. $\Re$ values mentioned below are computed in the kinematic phase, and vary with Pm by $\sim 10$–$15\%$ due to initial kinetic energy transients (see inset of Fig.\ref{fig: GR}).}

\begin{tabular}{c @{\hspace{0.45cm}} c @{\hspace{0.45cm}} c @{\hspace{0.55cm}} c @{\hspace{0.55cm}} c @{\hspace{0.45cm}} c @{\hspace{0.45cm}} c @{\hspace{0.45cm}}c @{\hspace{0.45cm}}c @{\hspace{0.45cm}} c @{\hspace{0.45cm}} c @{\hspace{0.3cm}} c}
\toprule
\mbox{Case} & $\textbf{B}_{\text{bc}}$ & $\Pm$   &   $\E$   & $\Ra$ &    $\Ra^*$   &      $\Re$    & $\Ro$ & $\Ro_{\ell}$ & $\bar{\ell}_{\text{v}}$   &   $P$ (ms) & $U (\si{\centi\meter\per\second})$ \\ 
\midrule
\mbox{\tbc{Baseline}} & \tbc{2} & \tbc{$2-35$} & \tbc{$1.66 \times 10^{-3}$} & \tbc{$5.8\times 10^{3}$}  & \tbc{$0.16$} & \tbc{42} & \tbc{$0.07$} & \tbc{0.15} & \tbc{6.73} & \tbc{$3.5$} & \tbc{$1.6\times 10^8$} \\
 \multicolumn{12}{c}{\hdashrule[0.5ex]{0.97\linewidth}{0.4pt}{2pt 2pt}} \\
  \mbox{Constant $\E$} & 2 & $2-35$ & $1.66 \times 10^{-3}$ & $1.3\times 10^{4}$ & $0.36$ & 66 & $0.11$ & 0.28 & 8.00 & $5.2$ & $2.5\times 10^8$\\
 & 2 & $2-35$ & $1.66 \times 10^{-3}$ & $3.0\times 10^{4}$ & $0.83$ & 91 & $0.15$ & 0.50 & 10.47 & $7.9$  & $3.4\times 10^8$\\
 \multicolumn{12}{c}{\hdashrule[0.5ex]{0.97\linewidth}{0.4pt}{2pt 2pt}} \\
  \mbox{Constant $\Ra^*$} & 2 & $2-35$ & $1.11 \times 10^{-3}$ & $1.3\times 10^{4}$ & $0.16$ & $69$ & $0.07$ & 0.15 & 6.73 & $3.5$ & $2.6\times 10^8$\\
 & 2 & $2-35$ & $0.73 \times 10^{-3}$ & $3.0\times 10^{4}$ & $0.16$ & $101$ & $0.07$ & $0.15$ & $6.73$ & $3.5$ & $3.8\times 10^8$\\
\addlinespace[3pt]
 \hline
 \addlinespace[3pt]
 & 4 & $2-35$ & $1.66 \times 10^{-3}$ & $5.8\times 10^{3}$ & $0.16$ & 42 & $0.07$ & $0.15$ & $6.73$ & $3.5$ & $1.6\times 10^8$\\ 
 & 4 & $3,4,6,10$ & $1.66 \times 10^{-3}$ & $1.3\times 10^{4}$ & $0.36$ & 66 & $0.11$ & $0.28$ & 8.00 & $5.2$ & $2.5\times 10^8$\\
 & 4 & $3,4,6,10$ & $1.66 \times 10^{-3}$ & $3.0\times 10^{4}$ & $0.83$ & 91 & $0.15$ & 0.50 & 10.47 & $7.9$ & $3.4\times 10^8$\\
 & 4 & $3,4,6,8$ & $1.11 \times 10^{-3}$ & $1.3\times 10^{4}$ & $0.16$ & 69 & $0.07$ & 0.15 & 6.73 & $3.5$ & $2.6\times 10^8$\\
 & 4 & $3,4,6,8$ & $0.73 \times 10^{-3}$ & $3.0\times 10^{4}$ & $0.16$ & 101 & $0.07$ & 0.15 & 6.73 & $3.5$ & $3.8\times 10^8$\\
\bottomrule
\end{tabular}
\label{tab: runs}
\end{table*}

\subsection{Output diagnostics} \label{sec: 2d}
\begin{figure}[h!]
\centering
\includegraphics[width=\linewidth]{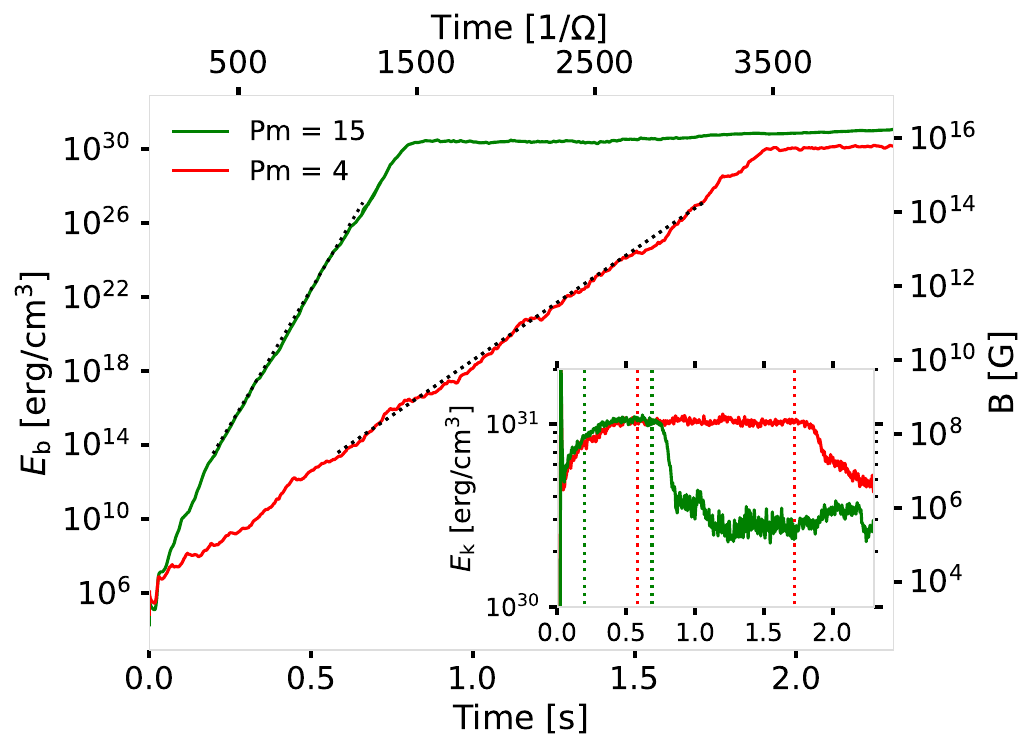}\hfill
\caption{Time evolution (in sec) of magnetic energy density $\Eb$ $(\si{\text{erg}\per\centi\cubic\meter})$ for $\Pm = 4$ (red) and $\Pm = 15$ (green) with $\Ra = 5.8 \times 10^3$ and $\E = 1.66 \times 10^{-3}$. The y-axis on the right is the magnetic field strength in Gauss, and the top x-axis denotes the rotation timescale. The black dotted lines represent the growth rate fit of the magnetic energy density in the range $\num{E13}-\SI{e27}{erg/cm^{3}}$. The inset shows the kinetic energy density $\Ek$ $(\si{\text{erg}\per\centi\cubic\meter})$ time series, with the vertical dotted lines indicating the region of the fits. We obtain $\gamma \text{ ($\Pm = 4)$} = 27.57 \pm 6.79 \text{ $\si{\per\second}$}$ and $\gamma \text{ ($\Pm = 15)$} = 67.53 \pm 6.10 \text{ $\si{\per\second}$}$.}
\label{fig: GR}
\end{figure} 
We now present the measure of the dynamo growth rate from the magnetic energy density time series output of the simulations. Figure \ref{fig: GR} shows the evolution of the magnetic energy density $\Eb$ $(\si{\text{erg}\per\centi\cubic\meter})$ as a function of time for $\Pm = 4$ (red) and $\Pm = 15$ (green) with $ \Ra = 5.8 \times 10^3$ and $\E=1.66 \times 10^{-3}$. The dotted black lines on the curves show the growth rate fit of the magnetic energy density in the kinematic phase. The inset shows the corresponding time evolution of the kinetic energy density $\Ek$ $(\si{\text{erg}\per\centi\cubic\meter})$, wherein the vertical dotted lines indicate the region used for the fits. The starting point of the growth rate estimation is chosen to exclude as much as possible the initial transients in both magnetic and kinetic energy densities. To omit non-linear effects from the Lorentz force, the endpoint is taken to be three orders of magnitude below the saturated magnetic energy density such that the ratio $\Eb/\Ek \lesssim 10^{-4}$. Therefore, considering the initial conditions of our simulations and the above constraints, we take the initial point of the fit to be the instant when the magnetic energy density first exceeds $10^{13}$ $\si{\text{erg}\per\centi\cubic\meter}$ and the endpoint as the stage when it reaches $10^{27}$ $\si{\text{erg}\per\centi\cubic\meter}$. This range has been chosen across all our simulations to calculate the growth rate. Variations of up to a few orders of magnitude around these values do not affect the results. To calculate the growth rate $\gamma$, we divide the data in the growth phase into 10 intervals. For each interval $i$, we find the best fit such that $\Eb(t) \propto \exp(\gamma_i t)$. We take the mean of all the $\gamma_i$ to be the growth rate $\gamma$ of the magnetic energy density for that simulation. An estimate of the corresponding error is quantified by the standard deviation of the $\gamma_i$ values relative to the mean $\gamma$.

\section{Results}\label{sec: res}
\subsection{Growth rate dependence on $\Pm$} \label{sec: growth}

\begin{figure}[h!]
     \centering
     \begin{subfigure}[b]{0.5\textwidth}
         \centering
         \includegraphics[width=\textwidth]{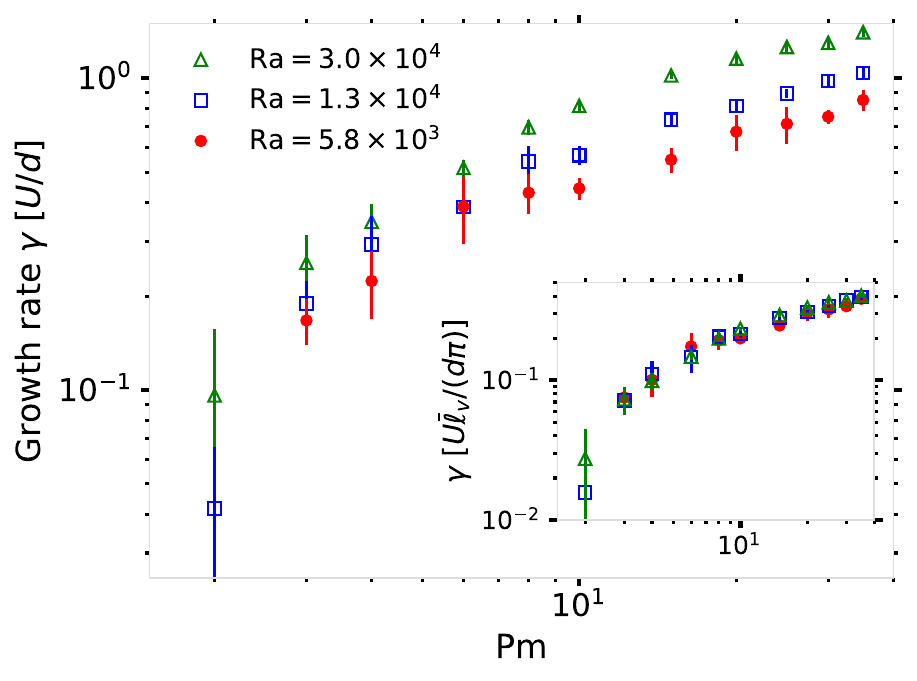}
         \caption{Simulation set with constant $\E = 1.66 \times 10^{-3}$.}
         \label{fig: GammaEkConstant}
     \end{subfigure}
     \hfill
     \begin{subfigure}[b]{0.5\textwidth}
         \centering
         \includegraphics[width=\textwidth]{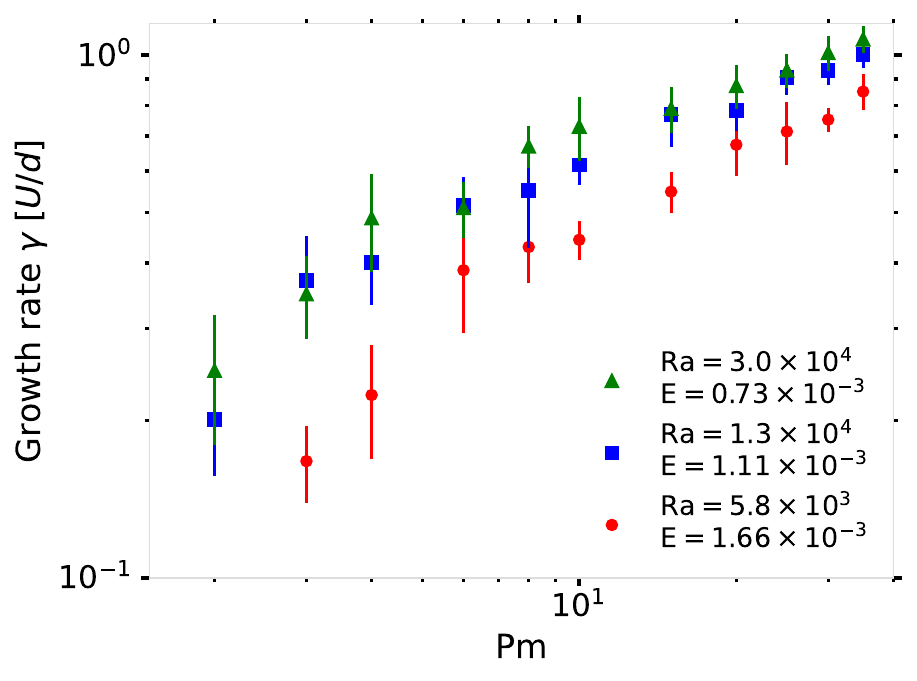}
         \caption{Simulation set  with constant $\Ra^* = 0.16$.}
         \label{fig: GammaRoConstant}
     \end{subfigure}
    \caption{Magnetic energy growth rate $\gamma$ (in units of $U/d$) as a function of Pm for three different values of (a) $\Ra$, and (b) $\Ra$ and $\E$. The filled red circle ({\color{red}{\rotatebox[origin=c]{45}{$\Large\pmb{\bullet}$}}}) is the baseline simulation set. The inset in (a) shows $\gamma$ in units of $U\bar{\ell}_{\text{v}}/(d\pi)$ as a function of Pm.}
\label{fig: Gamma}
\end{figure} 
Figure \ref{fig: Gamma} shows the variation of the growth rate $\gamma$ as a function of $\Pm$ wherein the baseline simulation set (first row in Table \ref{tab: runs}) is represented by filled red circles ({\color{red}{\rotatebox[origin=c]{45}{$\Large\pmb{\bullet}$}}}). The growth rate is normalised by the large-scale turnover rate $U/d$. In Figure \ref{fig: GammaEkConstant}, we keep $\E = 1.66 \times 10^{-3}$ and increase $\Ra$ and $\Pm$. We find that the growth rate increases as a function of $\Pm$ but smoothly transitions to a less steep dependence at high values of $\Pm$. A qualitatively similar behaviour is seen when $\Ra$ is increased. The inset in Fig. \ref{fig: GammaEkConstant} shows $\gamma$ in units of $U\bar{\ell}_{\text{v}}/(d\pi)$. A collapse occurs for all three $\Ra$ cases on the same curve as a function of Pm, indicating that for the constant $\E$ case, $d\pi/(U\bar{\ell}_{\text{v}}$) is the relevant timescale for the growth of the instability. In Fig.~\ref{fig: GammaRoConstant}, we keep $\Ra^* = 0.16$ and decrease $\E$ while increasing $\Ra$ and $\Pm$. Again, $\gamma$ is found to increase with $\Pm$, becoming less steep at high $\Pm$, with the overall trend remaining unchanged with increasing $\Ra$. When comparing the magnitude of growth rates in Figs. \ref{fig: GammaEkConstant} and \ref{fig: GammaRoConstant}, stronger variations are found for the constant $\E$ than for the constant $\Ra^*$ case. For both sets, though, we see that the growth rate has not yet saturated at high $\Pm$. The 3D global simulations at even higher values of Pm are computationally costly. Hence, later in Section \ref{sec: KK}, we employ a theoretical model, the Kazantsev dynamo model, motivated in Section \ref{sec: magnetic field topology}, to understand the behaviour of growth rates as we approach the large $\Pm$ limit relevant to PNS. 

\begin{figure}[h!]
\centering
\includegraphics[width=\linewidth]{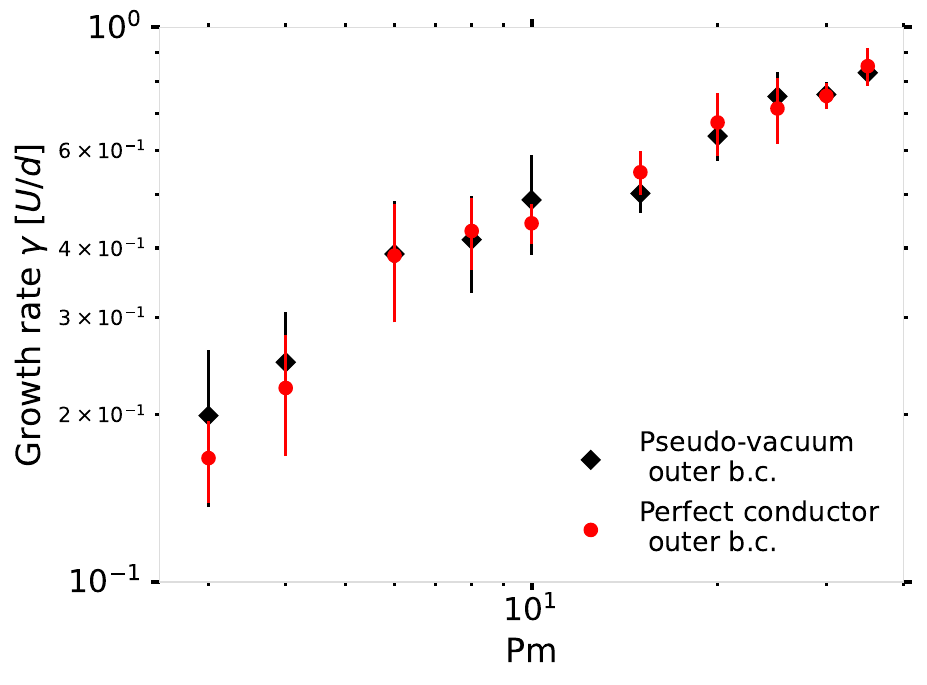}\hfill
\caption{Magnetic energy growth rate $\gamma$ (in units of $U/d$) as a function of Pm, for pseudo-vacuum (\blackdiamondshape\mbox{}) and perfect conductor ({\color{red}{\rotatebox[origin=c]{45}{$\Large\pmb{\bullet}$}}}) outer magnetic boundary conditions, with $\Ra=5.8 \times 10^3 \mbox{ and } \E = 1.66 \times 10^{-3}$.} 
\label{fig: BC}
\end{figure}

We also look at the effect of the magnetic boundary condition on the growth rate of the convective dynamo. Figure \ref{fig: BC} shows $\gamma$ (in units of $U/d$) as a function of $\Pm$ for different outer magnetic boundary conditions. Again, the filled red circles represent the baseline simulation set. As seen from the figure, $\gamma$ is independent of the choice of outer magnetic boundary condition, which is expected in the kinematic phase and is consistent with the predictions of \cite{favier2013growth}. Although results are presented only for $\Ra  = 5.8 \times 10^3$ and $\E = 1.66 \times 10^{-3}$, selected additional simulations (not shown) with varying $\Ra$ and $\E$ confirm this result.

\subsection{Topology and length scales of magnetic field} \label{sec: magnetic field topology}

\begin{figure*}
     \centering
     \begin{subfigure}[b]{0.49\textwidth}
         \centering
         \includegraphics[width=\textwidth, trim={2cm 2.2cm 1cm 5.1cm}, clip]{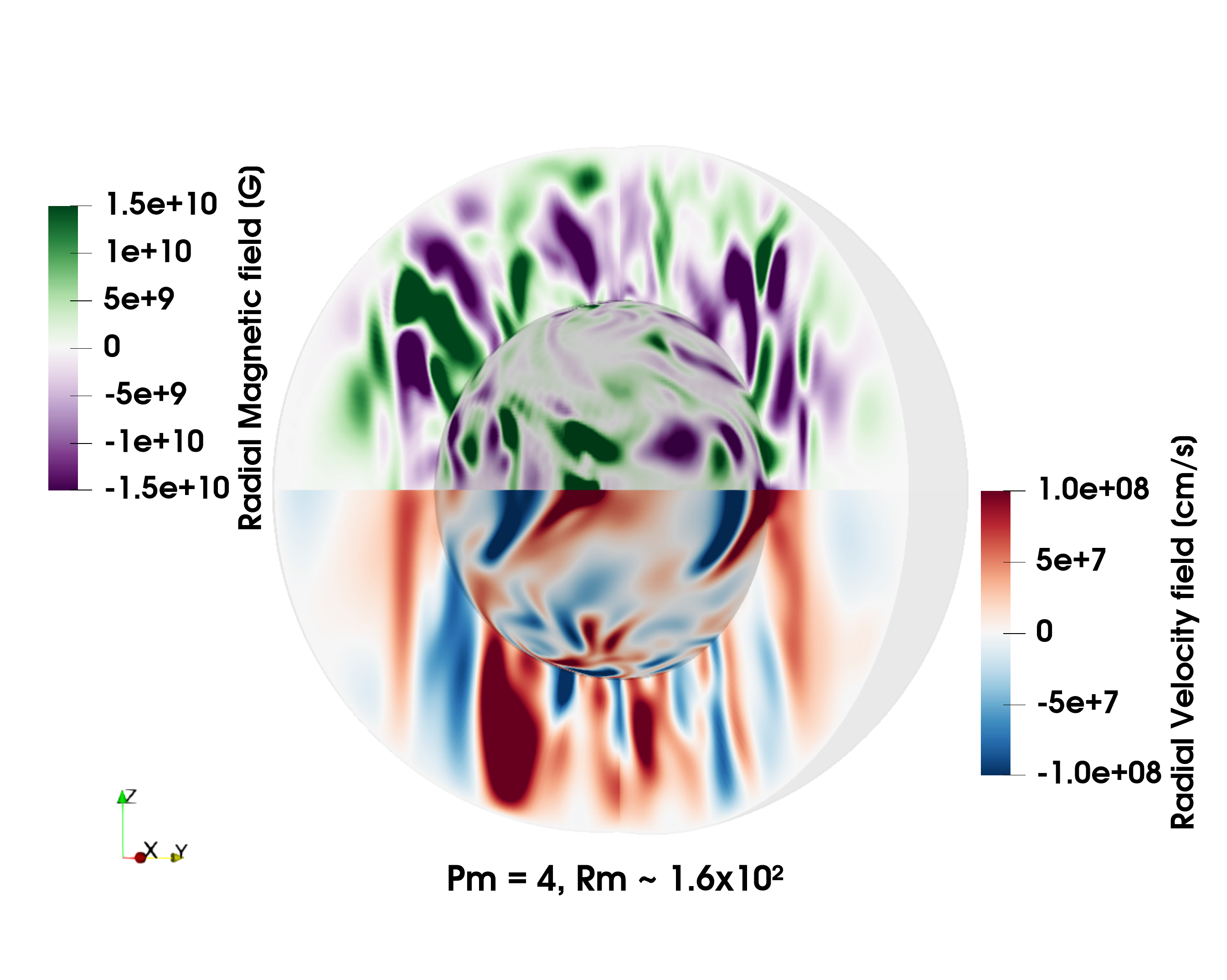}
         \caption{}
         \label{fig: RadialPm_4}
     \end{subfigure}
     \hfill
     \begin{subfigure}[b]{0.49\textwidth}
         \centering
         \includegraphics[width=\textwidth, trim={2cm 2.2cm 1cm 5.1cm}, clip]{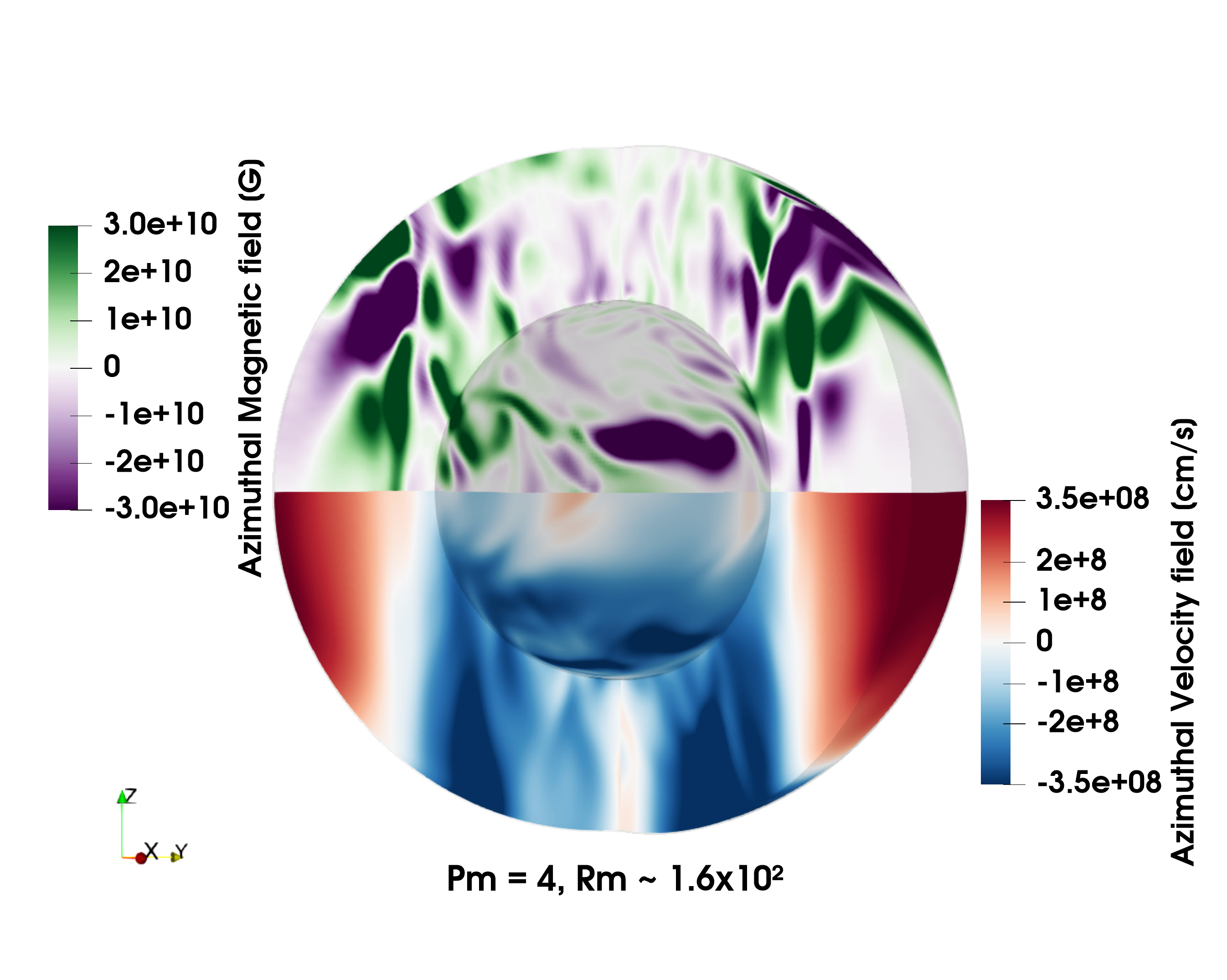}
         \caption{}
         \label{fig: ZonalPm_4}
     \end{subfigure}
     \hfill
     \begin{subfigure}[b]{0.49\textwidth}
         \centering
         \includegraphics[width=\textwidth, trim={2cm 2.2cm 1cm 5.1cm}, clip]{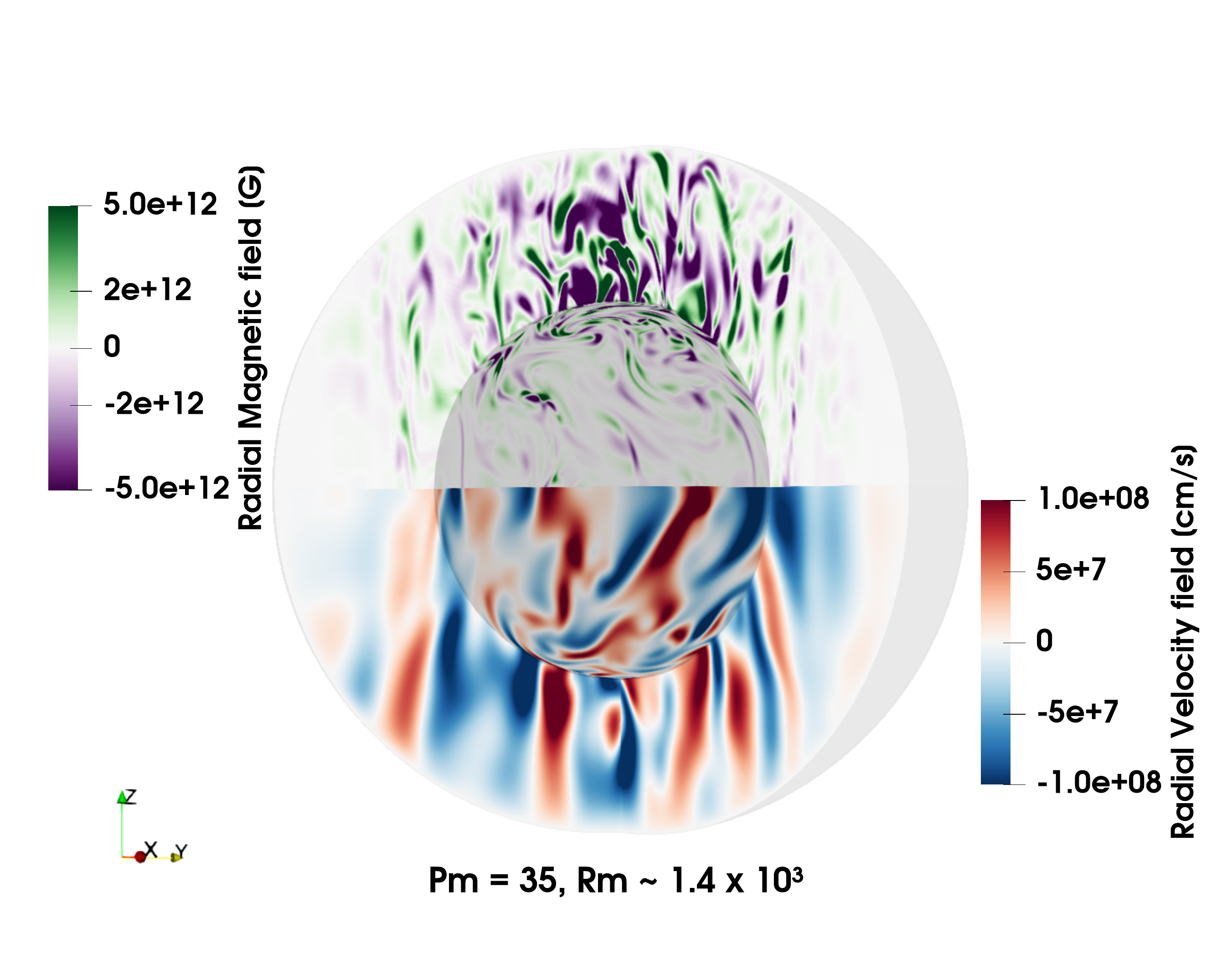}
         \caption{}
         \label{fig: RadialPm_35}
     \end{subfigure}
     \hfill
     \begin{subfigure}[b]{0.49\textwidth}
         \centering
         \includegraphics[width=\textwidth, trim={2cm 2.2cm 1cm 5.1cm}, clip]{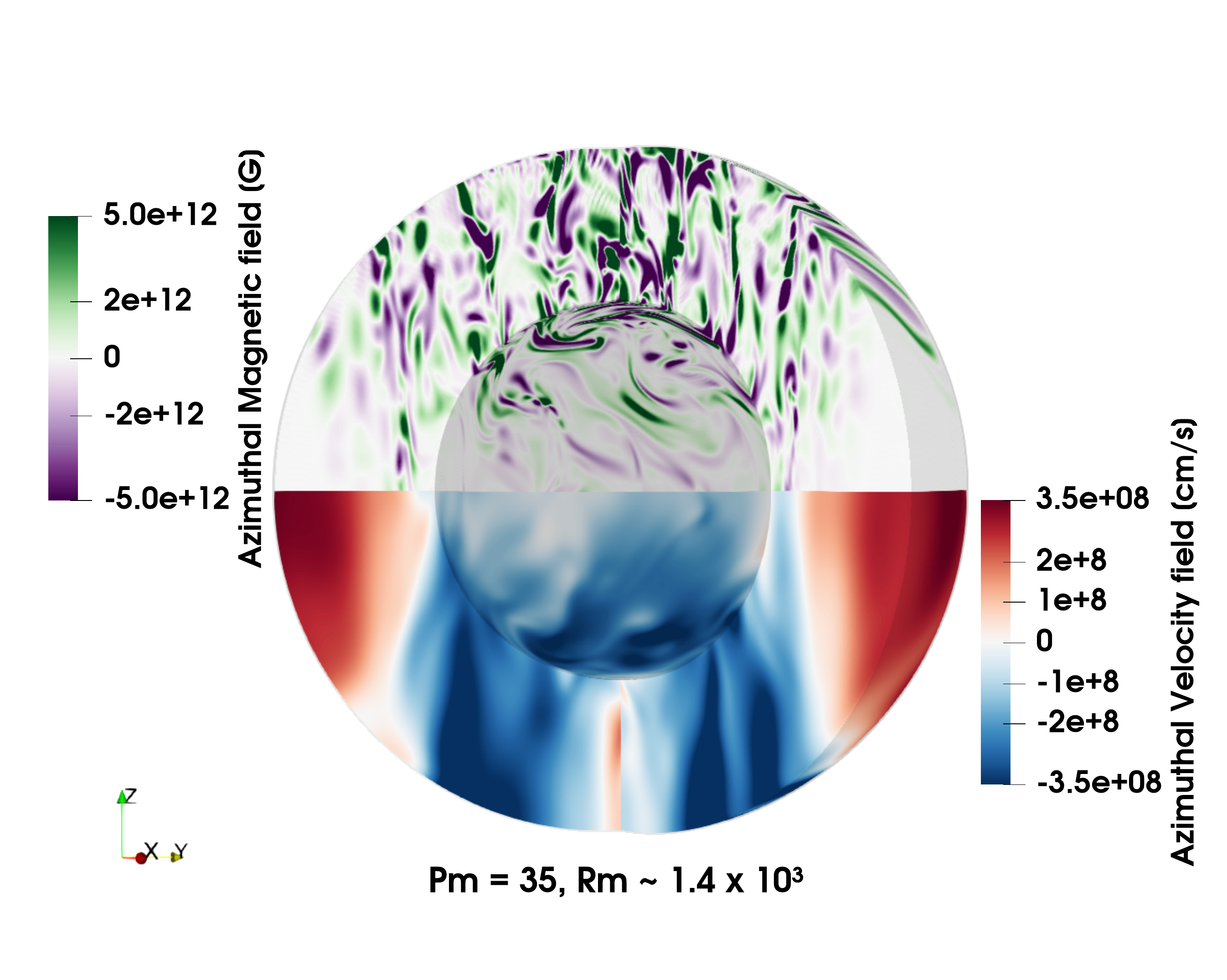}
         \caption{}
         \label{fig: ZonalPm_35}
     \end{subfigure}
\caption{3D rendering of (a), (c) the radial component, and (b), (d) the azimuthal component of the magnetic field in the {\it Northern hemisphere} and the velocity field in the {\it Southern hemisphere} during the kinematic phase. The top panels correspond to low $\Pm \mbox{ } (=4)$, $\Rm\mbox{ } (\sim 1.6 \times 10^2)$ and bottom panels to high $\Pm \mbox{ }(=35)$, $\Rm\mbox{ } (\sim 1.4 \times 10^3)$. In both simulations, we have $\Ra=5.8 \times 10^3$ and $\E=1.66 \times 10^{-3}$. The inner spherical surface is at $r=1.1r_i$, and the outer surface is at $r=r_o$.}
\label{fig: 3D}
\end{figure*}

Next, we study the topology and the characteristic length scale of the amplified magnetic field. To visualise the length scale of the fields, we show in Figure \ref{fig: 3D} 3D renderings of the radial component (Figs. \ref{fig: RadialPm_4}, \ref{fig: RadialPm_35}) and the azimuthal component (Figs. \ref{fig: ZonalPm_4}, \ref{fig: ZonalPm_35}) of the magnetic field in the {\it Northern hemisphere} and the velocity field in the {\it Southern hemisphere}, during the kinematic phase. The top panels (Figs. \ref{fig: RadialPm_4}, \ref{fig: ZonalPm_4}) correspond to a low value of $\Pm = 4$ while the bottom panels (Figs. \ref{fig: RadialPm_35}, \ref{fig: ZonalPm_35}) represent a high value, $\Pm=35$. In both simulations, we have fixed $\Ra$ = $5.8 \times 10^3$ and $\E$ = $1.66 \times 10^{-3}$. As $\Pm$ is increased, we find that the radial magnetic field length scale decreases and becomes dominated by small-scale structures, as seen when comparing Figs. \ref{fig: RadialPm_4} and \ref{fig: RadialPm_35}. On the other hand, at both low and high $\Pm$, the radial velocity field exhibits the characteristic pattern of convection, with the Taylor columns stretching over the entire spherical shell. This indicates a scale separation between the magnetic and velocity fields as $\Pm$ is increased. The scale separation is also discernible when comparing the azimuthal components of the fields, depicted in Figs. \ref{fig: ZonalPm_4} and \ref{fig: ZonalPm_35}. A large-scale prograde zonal velocity field is observed, independently of $\Pm$, as expected in the kinematic phase. By contrast, we find large-scale structures of the magnetic field at low $\Pm$ (Fig.~\ref{fig: ZonalPm_4}) and as we increase $\Pm$, small-scale structures dominate (Fig.~\ref{fig: ZonalPm_35}). Additionally, we observe that at high $\Pm$, most of the magnetic field amplification is concentrated within the tangent cylinder (Figs. \ref{fig: RadialPm_35} and \ref{fig: ZonalPm_35}), whereas at low $\Pm$ the dominant structures are found outside (Figs \ref{fig: RadialPm_4} and \ref{fig: ZonalPm_4}). 
The radial and azimuthal components of the field exhibit similar behaviour when $\Ra$ is increased.

\begin{figure}[h!]
     \centering
     \begin{subfigure}[b]{0.5\textwidth}
         \centering
         \includegraphics[width=\textwidth]{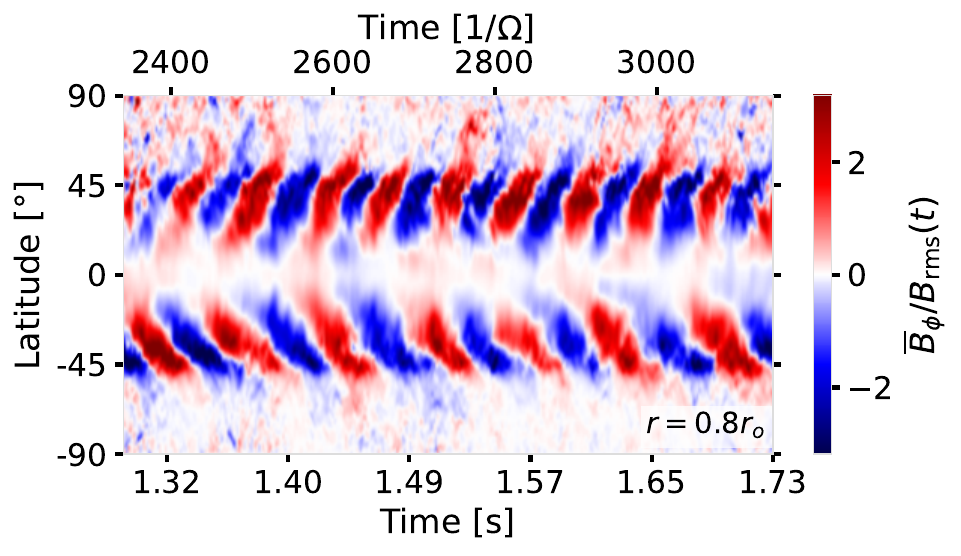}
         \caption{$\Ra = 5.8 \times 10^{3}$, $\Pm =4$ ($\Rm \sim 1.6 \times 10^{2}$).}
         \label{fig: Oscillationpm4}
     \end{subfigure}
     \hfill
     \begin{subfigure}[b]{0.5\textwidth}
         \centering
         \includegraphics[width=\textwidth]{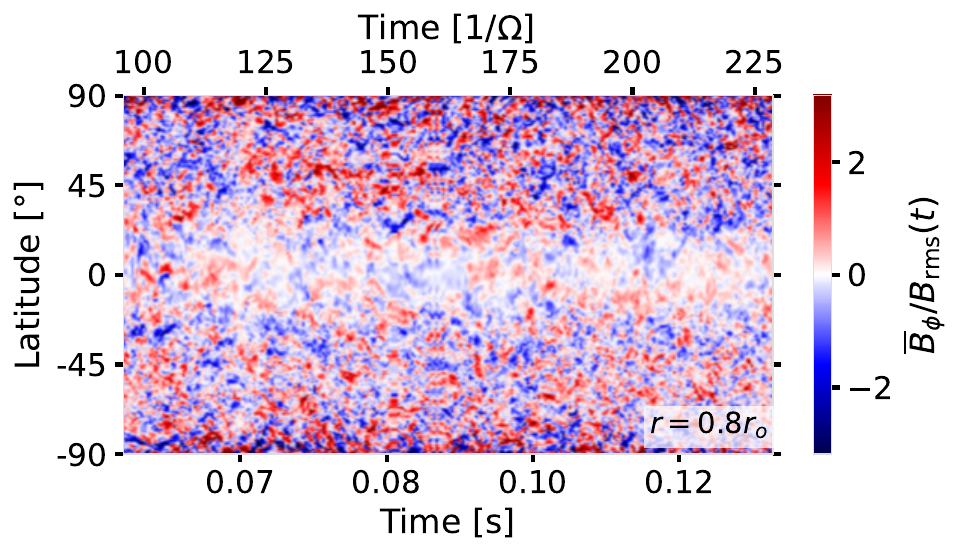}
         \caption{$\Ra = 3.0 \times 10^{4}$, $\Pm =35$ ($\Rm \sim 2.9 \times 10^3$).}
         \label{fig: Oscillationpm35}
     \end{subfigure}
    \caption{Time evolution of $\overline{B}_{\phi}/B_{\text{rms}}(t)$ - the $\phi$ average of the azimuthal component of the magnetic field normalised by its RMS value at each time step - during the kinematic phase. For both, $\E = 1.66 \times 10^{-3}$.}
\label{fig: Oscillation}
\end{figure}

Now, to probe the mechanism responsible for the magnetic field amplification, we look at the butterfly diagrams, which show the $\phi$-averaged azimuthal component of the magnetic field at $r=0.8r_o$, denoted by $\overline{B}_{\phi}$, as a function of time and colatitude $\theta$. To compensate for the amplitude growth, we normalise $\overline{B}_{\phi}$ by its RMS value at each time step, given by,
\begin{equation}
B_{\mathrm{rms}}(t) = \sqrt{\frac{1}{N_{\theta}}\sum_{i=1}^{N_{\theta}}(\overline{B}_{\phi,i}(t))^2}
\end{equation}
where $N_{\theta}$ is the latitudinal resolution. Figure \ref{fig: Oscillation} shows the time evolution of $\overline{B}_{\phi}/B_{\text{rms}}(t)$ in the growth phase for (a) low $\Rm$ and (b) high $\Rm$. We observe a pronounced large-scale dynamo wave, typical of a mean field $\alpha\Omega$ dynamo for $\Rm \lesssim 10^3$, as shown in Fig. \ref{fig: Oscillationpm4}, where a coherent magnetic cycle with a period $\sim \SI{60}{ms}$ is visible. These periodic reversals are completely lost when $\Rm\gtrsim 2\times 10^3$ and the magnetic field is dominated by small-scale structures (Fig.~\ref{fig: Oscillationpm35}). Thus, from Figs. \ref{fig: 3D} and \ref{fig: Oscillation}, we infer a gradual transition from a large-scale to a small-scale dynamo as $\Pm$ and $\Ra$ are increased. It is to be noted that this transition is governed by the underlying flow structure \cite{richardson1926atmospheric}, through $\Ro$ and $\Ra$.

\begin{figure}[h!]
 \centering
 \includegraphics[width=0.5\textwidth]{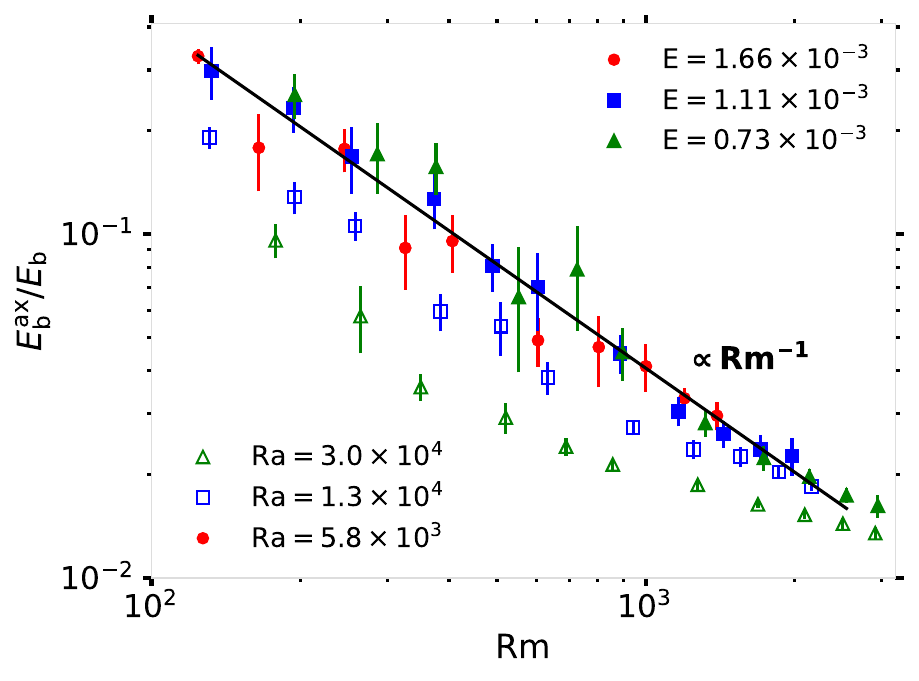}
\caption{Axisymmetric to total magnetic energy density ratio $\Ebax/\Eb$ as a function of $\Rm$, for both sets of simulations. As in Fig. \ref{fig: Gamma}, symbols represent the constant $\E$ (empty symbols) and constant $\Ra^*$ (filled symbols) simulations, with {\color{red}{\rotatebox[origin=c]{45}{$\Large\pmb{\bullet}$}}} as baseline simulation set. The solid line represents the best-fit scaling $\Ebax/\Eb = 42.05 \mbox{ }\Rm^{-1}$.}
\label{fig: EabyEb}
\end{figure}
\begin{figure}[h!]
 \centering
 \includegraphics[width=0.5\textwidth]{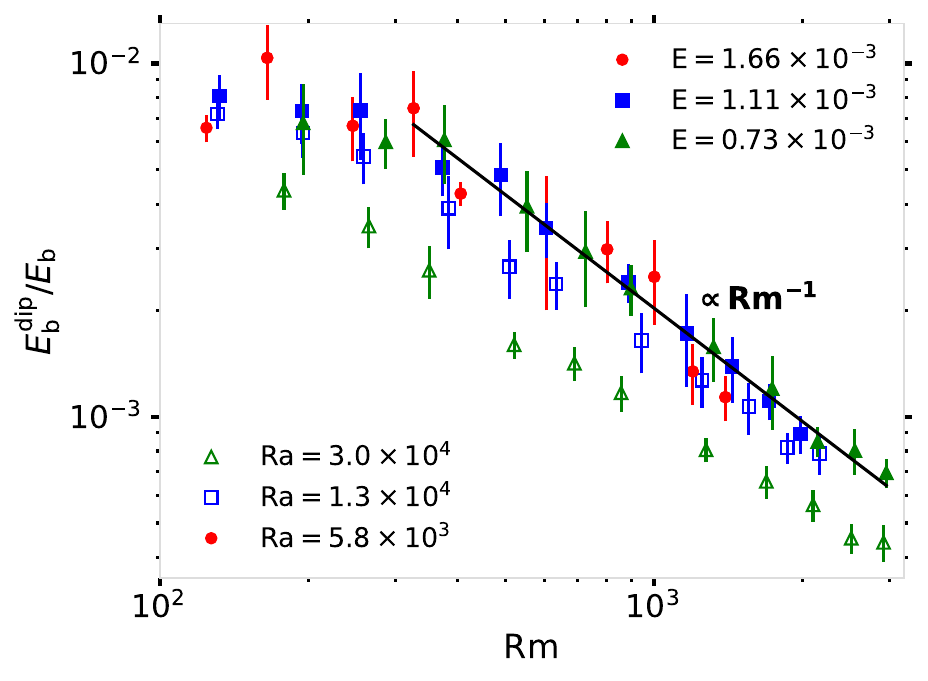}
\caption{Dipolar to total magnetic energy density ratio $\Ebdip/\Eb$ as a function of $\Rm$, for both sets of simulations. As in Fig. \ref{fig: Gamma}, the symbols represent the constant $\E$ (empty symbols) and constant $\Ra^*$ (filled symbols) simulations, with {\color{red}{\rotatebox[origin=c]{45}{$\Large\pmb{\bullet}$}}} as baseline simulation set. The solid line represents the best-fit scaling $\Ebdip/\Eb = 3.25 \mbox{ }\Rm^{-1}$.}
\label{fig: EdbyEb}
\end{figure}

Next, we study the topology of the generated magnetic field. We denote the axisymmetric magnetic energy density by $\Ebax$ and the dipolar magnetic energy density by $\Ebdip$. The fractions $\Ebax/\Eb$ and $\Ebdip/\Eb$ provide a measure of the symmetry of the growing mode and the relative strength of the large-scale field, respectively. Figure~\ref{fig: EabyEb} shows $\Ebax/\Eb$ as a function of $\Rm$ for both constant $\E$ (empty symbols) and constant $\Ra^*$ (filled symbols) simulations, indicating a decrease in $\Ebax/\Eb$ with increasing $\Rm$. The black solid line shows the best-fit scaling $\Ebax/\Eb \propto \Rm^{-1}$, which applies independently of $\E$ for the constant $\Ra^*$ simulations. On the other hand, for the constant $\E$ case, the decay in $\Ebax/\Eb$ is comparatively steeper at low $\Rm$ but becomes less steep at high $\Rm$, especially for low values of $\E$. Hence, in both sets of simulations, the most unstable mode of the magnetic field becomes increasingly non-axisymmetric as one approaches high $\Rm$ values. 

Similarly, the dipolarity displayed in Fig.~\ref{fig: EdbyEb} shows a decrease in $\Ebdip/\Eb$ as a function of $\Rm$. Again, the black solid line is the best-fit scaling $\Ebdip/\Eb \propto \Rm^{-1}$ for the constant $\Ra^*$ simulations, valid for $\Rm \gtrsim 300$. While at lower $\Rm$, the dipolarity remains constant. The constant $\E$ case is also seen to follow the $\Rm^{-1}$ scaling, but with a normalisation factor that depends on $\Ra$. Hence, from Figs. \ref{fig: EabyEb} and \ref{fig: EdbyEb}, we deduce that the magnetic field becomes highly non-axisymmetric and multipolar at very small diffusivities. If this scaling can be extrapolated to PNS conditions, where $\Rm\sim10^{17}$ \cite{lander2021generating}, one therefore expects that only a minute fraction of the magnetic energy resides in the large-scale component by the end of the kinematic phase. For a total magnetic field of order $\sim \SI{e16}{G}$, the corresponding dipolar strength is merely $\sim \SI{e8}{G}$, which is far smaller than the $\SI{e15}{G}$ inferred for magnetars. If the large-scale magnetic field of magnetars originates from a convective dynamo, one possible scenario is that it undergoes further amplification during the nonlinear phase, similar to that reported by Raynaud et al. \cite{raynaud2020magnetar}.

\begin{figure}[h!]
     \centering
     \includegraphics[width=0.5\textwidth]{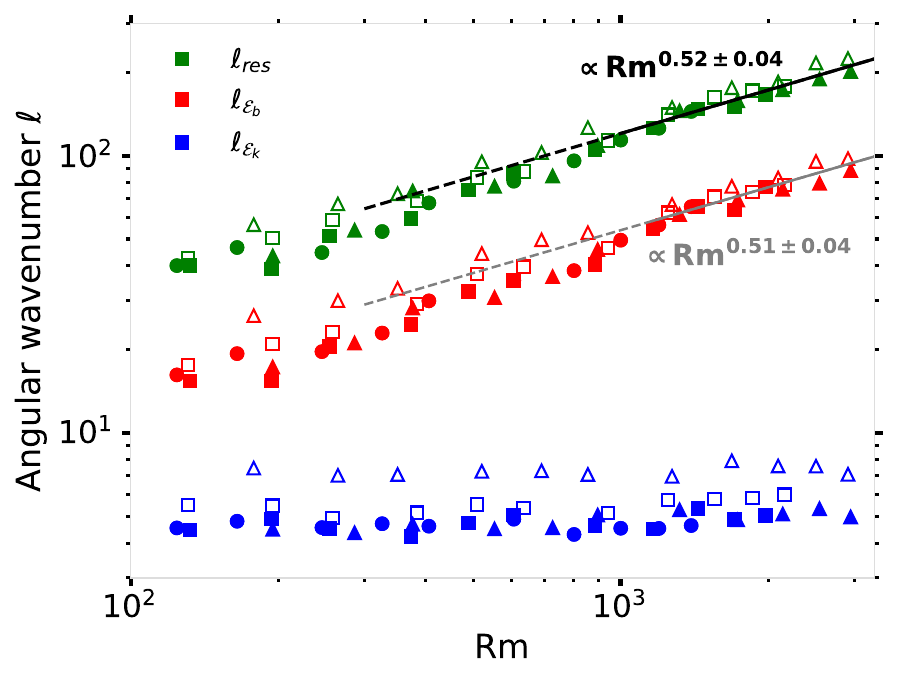}
\caption{Characteristic angular wavenumber for kinetic energy $\ell_{\mathcal{E}_k}$ (blue), magnetic energy $\ell_{\mathcal{E}_b}$ (red) and magnetic dissipation energy $\ell_{res}$ (green) as a function of $\Rm$. As in Fig. \ref{fig: Gamma}, the symbols represent the constant $\E$ (empty symbols) and constant $\Ra^*$ (filled symbols) simulations. The solid lines represent the best fit scaling $\ell_{\mathcal{E}_b} = 1.55 \mbox{ } \Rm^{0.51}$ (grey) and $\ell_{res} = 3.32 \mbox{ }\Rm^{0.52}$ (black). The dashed lines are an extrapolation of the fits.}
\label{fig: wavenumber}
\end{figure} 

To quantify the length scales of the fields, we introduce the characteristic angular wavenumber for kinetic energy, magnetic energy, and magnetic dissipation energy, denoted by $\ell_{\mathcal{E}_k}$, $\ell_{\mathcal{E}_b}$, and $\ell_{res}$, respectively, as
\begin{equation}
    \ell_{\mathcal{E}_k} = \dfrac{\sum \ell \mathcal{E}_k }{\sum \mathcal{E}_k}; \mbox{ } \ell_{\mathcal{E}_b} = \dfrac{\sum \ell \mathcal{E}_b }{\sum \mathcal{E}_b}; \mbox{ } \ell_{res} = \dfrac{\sum \ell^2(\ell+1)\mathcal{E}_b }{\sum \ell(\ell+1)\mathcal{E}_b}
\end{equation}
Here, $\mathcal{E}_k$ and $\mathcal{E}_b$ represent the kinetic energy and magnetic energy, respectively, in the angular wavenumber $\ell$ at an instant during the kinematic phase. The summation $\sum$ is over $\ell \in [0,\mbox{ }\ell_{\text{max}}]$. Figure \ref{fig: wavenumber} shows the angular wavenumber $\ell_{\mathcal{E}_k}$ (blue), $\ell_{\mathcal{E}_b}$ (red) and $\ell_{res}$ (green) as a function of $\Rm$ for constant $\E$ and constant $\Ra^*$ simulations. As expected for kinematic simulations, $\ell_{\mathcal{E}_k}$ remains approximately independent of $\Rm$, whereas $\ell_{\mathcal{E}_b}$ and $\ell_{res}$ increase with $\Rm$ and converge to a $ \sim \sqrt{\Rm}$ scaling for $\Rm \gtrsim 10^3$. The increasing difference between $\ell_{\mathcal{E}_b}$ and $\ell_{\mathcal{E}_k}$ further confirms the scale separation between the magnetic and the velocity fields observed in Fig.~\ref{fig: 3D}. The scaling for $\ell_{res}$ can be understood by equating the shear production rate with the resistive decay rate in the induction equation,
\begin{equation}
  \frac{U}{d} \sim
  \frac{\eta_{r_o}}{(d/\ell_{res})^2},
  \;\;\;\;\;
  \ell_{res}^2 \sim \frac{U d}{\eta_{r_o}} = \Rm,
\end{equation}
giving \cite{guilet2022mri,schekochihin2004simulations},
\begin{equation}
  \ell_{\mathrm{res}} \;\sim\; \sqrt{\Rm}.
\end{equation}
This suggests that, in PNSs at $\Rm \sim 10^{17}$, magnetic field amplification will occur predominantly at very large angular wavenumbers, of order $\sim 10^8$, strongly supporting a small-scale dynamo.

\section{Large $\Pm$ Limit}\label{sec: KK}
Due to computational constraints, direct numerical simulations (DNS) at higher values of $\Pm$ and $\Ra$ are not feasible for the system under study. Hence, we seek a theoretical framework to understand the mechanism of magnetic field generation and extrapolate it to the large $\Pm$ regime relevant to PNSs. Our DNS results show that, with increasing $\Pm$ and $\Ra$, the magnetic field becomes localised at small scales, and the large-scale dynamo wave is lost. This motivates us to employ a small-scale dynamo model, which describes the amplification of the magnetic field in a turbulent conducting fluid \cite{batchelor1950spontaneous,schekochihin2001structure}. We therefore use the canonical analytical formalism developed by Kazantsev \cite{kazantsev1968enhancement} wherein a random turbulent velocity field is shown to enhance the initially weak magnetic field in the kinematic phase. We follow Schober et al. (2012) \cite{schober2012magnetic} to model the statistical properties of this random flow. A brief introduction to the chosen approach is given in Section \ref{sec: KK1}. The statistical properties of the flow incorporate parameters constrained using the DNS velocity field, thereby linking the simulations to the theoretical model. The estimation of these parameters is discussed in Section~\ref{sec: KK2}. Within this framework, we solve the Kazantsev model, determine the dependence of the magnetic energy growth rate on $\Pm$ \cite{bovino2013turbulent}, and compare the theoretical predictions with the numerical results in Section \ref{sec: KK3}.

\subsection{The Kazantsev dynamo model}\label{sec: KK1}

The canonical model assumes a statistically homogeneous, isotropic and random Gaussian velocity field with zero mean that is $\delta$-correlated in time. Under these assumptions, the correlation tensor can be written as, 
\begin{equation}
    \langle  {v}_i(\vec{r}_1, t) \,  {v}_j(\vec{r}_2, t') \rangle = T_{ij}(r) \, \delta(t - t')
\end{equation}
where, ${v}_i$ and ${v}_j$ denote the $i^{th}$ and $j^{th}$ turbulent components of the velocity field at position $\vec{r}_1$ and $\vec{r}_2$ at time $t$ and $t'$, respectively, $\langle\mbox{ }\rangle$ represents the statistical average, $T_{ij}(r)$ is the two-point correlation function and ${r}=\left|\vec{r}_1-\vec{r}_2\right|$ is the distance between the two positions. Assuming isotropy in small-scale turbulence, $T_{ij}(r)$ can be decomposed into a longitudinal $T_L(r)$ and a transverse $T_N(r)$ component such that \cite{batchelor1950spontaneous},
\begin{equation}
T_{ij}(r) = \frac{r_i r_j}{r^2}T_L(r) + \left(\delta_{ij}-\frac{r_i r_j}{r^2} \right)T_N(r).
\end{equation}
Similarly, considering the magnetic field to be homogeneous, isotropic and random Gaussian with zero mean, a two-point correlation function $M_{ij}(r,t)$ can be decomposed into a longitudinal $M_L(r,t)$ and a transverse $M_N(r,t)$ component. Now, using the induction equation and the solenoidal condition, the time evolution of the magnetic field fluctuations $M_L(r,t)$ can be derived. Assuming the separation of space and time variables of the form, 
\begin{equation}
   M_L(r, t)=\dfrac{1}{(r^2\sqrt{\kappa_{\text{d}}})}\psi(r)e^{2 \Gamma t},
\end{equation}
where $2\Gamma$ is the corresponding growth rate of the magnetic correlation function, the time evolution of $M_L(r,t)$ reduces to an eigenvalue problem for $\psi$ described by the Kazantsev equation \cite{kazantsev1968enhancement,schober2012magnetic},
\begin{equation}
    - \kappa_{\text{d}}(r) \psi''(r) + V(r) \psi(r) = - \Gamma \psi(r).
    \label{kaz}
\end{equation}
Here, prime denotes derivative with respect to $r$. The effective potential $V(r)$ and effective diffusion of magnetic correlation $\kappa_{\text{d}}(r)$ are, respectively, given by,
\begin{equation}
    V(r) = \frac{\kappa_{\text{d}}''(r)}{2} - \frac{(\kappa_{\text{d}}'(r))^2}{4 \kappa_{\text{d}}(r)} + \frac{2 T_N'}{r} + \frac{2 (T_L- T_N + \kappa_{\text{d}}(r))}{r^2}
\end{equation}
\begin{equation}\label{eqn: etat}
    \kappa_{\text{d}}(r) = \eta +T_L(0)-T_L(r) = \eta + \eta_t(r),
\end{equation}
where $\eta$ is the magnetic diffusivity and $T_L(0)-T_L(r)$ can be interpreted as the turbulent diffusivity $\eta_t$.

Solving equation \eqref{kaz} requires a model for the two-point velocity correlation functions $T_L(r)$ and $T_N(r)$, which follows from Richardson's law of turbulent diffusion. At some time $t$, consider two Lagrangian particles at position $\vec{r}_1$ and $\vec{r}_2 $ with velocity $\vec{v}_1$ and $\vec{v}_2$, respectively, such that the separation distance $|\vec{r}_1-\vec{r}_2| = {r}$ and the velocity difference $|\vec{v}_1-\vec{v}_2| =\Delta{v}({r})$, where $|\mbox{ } |$ represents the magnitude of the vector. According to Richardson's law \cite{richardson1926atmospheric}, in the inertial range, the characteristic velocity difference across a separation $r$ follows,
\begin{equation}
\Delta{v}({r}) \propto r^{\theta},
\end{equation}
where $\theta$ is an exponent. Using this relation and dimensional analysis, one can deduce $\eta_t$ such that,
\begin{equation}
    \eta_t \sim r\Delta v(r)  \sim rr^{\theta} \sim r^{1+\theta}.
 \end{equation}
This implies that, in the inertial range, the longitudinal and transverse velocity components can be written as,
\begin{equation}\label{eqn delv}
    \Delta{v}_L({r}) \sim r^{\theta_L}; \mbox{ } \Delta{v}_N({r}) \sim r^{\theta_N}.
\end{equation}
From these, the two-point correlations are obtained as
\begin{equation}\label{eqn gtltn}
    T_L({r}) \sim r^{1+\theta_L}; \mbox{ } T_N({r}) \sim r^{1+\theta_N}.
\end{equation}
In this framework, following \cite{schober2012magnetic}, a general longitudinal correlation function for different length scales is 
\begin{equation}
T_L(r) =
\begin{cases} 
\frac{UD}{3} \left[ 1 - \text{Re}^{(1-\theta_L)/(1+\theta_L)} \left( \frac{r}{D} \right)^2 \right], & 0 < r < \ell_c, \\[10pt]
\frac{UD}{3} \left[ 1 - \left( \frac{r}{D} \right)^{ 1 + \theta_L} \right], & \ell_c < r < D, \\[10pt]
0, & D < r.
\end{cases}
\label{tl}
\end{equation}
Here $(0, \ell_c)$ denotes the diffusive range, where $\ell_c = D{\Re}^{-1/(1+\theta_L)}$ is the cutoff length scale of the turbulence, with $D$ as the forcing scale, while the interval $(\ell_c, D)$ corresponds to the inertial range. A similar expression for the transverse correlation function holds \cite{schober2012magnetic}
\begin{equation}
T_N(r) =
\begin{cases} 
\frac{UD}{3} \left[ 1 - \zeta\,\text{Re}^{(1-\theta_N)/(1+\theta_N)} \left( \frac{r}{D} \right)^2 \right], & 0 < r < \ell_c, \\[10pt]
\frac{UD}{3} \left[ 1 - \zeta \left( \frac{r}{D} \right)^{ 1+\theta_N} \right], & \ell_c < r < D, \\[10pt]
0, & D < r.
\end{cases}
\label{tn}
\end{equation}
The parameter $\zeta$ quantifies the relative amplitude of the transverse and longitudinal velocity fluctuations. 

Under the assumption of incompressibility in Kolmogorov turbulence, $\zeta = 1.67$ and $\theta_L=\theta_N=0.33$ \cite{schober2012magnetic}. 
In our model, however, we allow the exponents to differ ($\theta_L \neq \theta_N$) to account for deviations from isotropy in rapidly rotating convective turbulence as well as the effects of compressibility. Similar modifications have been studied while modelling anisotropic turbulence \cite{Romano_Antonia_2001} and it is important given the role of anisotropy in dynamo processes, as demonstrated in previous studies \cite{vishniac2014properties,kolokolov2011magnetic,singh2017enhancement}. In general, $\theta_L$, $\theta_N$, and $\zeta$ are regarded as free parameters that are determined from numerical or experimental data. Therefore, using the DNS output velocity field together with equation \eqref{eqn delv}, we now constrain these parameters.

\subsection{DNS based parameter estimation}\label{sec: KK2}
We take the output DNS velocity field during the kinematic phase once the hydrodynamics have reached a statistically steady state. For any two position vectors $\vec{r}_1$ and $\vec{r}_2$, recall that their separation distance is $|\vec{r}_1-\vec{r}_2| = {r}$, implying $\vec{r}_1-\vec{r}_2 = \vec{r}$. As equation \eqref{eqn delv} is valid only within the inertial range, we define this range by assuming that a scaling close to the Kolmogorov holds. The lower limit is set to exclude the dissipation regime whose extent is determined by the dissipative scale $\sim D\Re^ {-3/4}$, while the upper limit is chosen to omit the forcing regime. Taking the forcing scale $D$ to be the shell width $d$, we demarcate the inertial range as
\begin{equation}
    r \in \left[\dfrac{3}{2}\dfrac{d}{Re^{3/4}}, \dfrac{d}{2}\right].
\end{equation}
The numerical prefactors are chosen to exclude regions where the inertial regime overlaps with the other two regimes. In order to encompass the inertial range across all our simulations, we choose to calculate pairwise separations within $r\in [0.02d,1.3d]$. For each $r$, we decompose the DNS output velocity field at the two positions, $\vec{v}_1$ and $\vec{v}_2$, into their longitudinal $\vec{v}_L$ and transverse $\vec{v}_N$ components. The longitudinal component is collinear with the separation vector $\vec{r}$ while the transverse component is orthogonal to it, such that,
\begin{equation}
    {v}_{1_L} =\vec{v}_1 \cdot \dfrac{\vec{r}}{r}; \mbox{ } \mbox{ } \mbox{ } {v}_{2_L} =\vec{v}_2 \cdot \dfrac{\vec{r}}{r},
\end{equation} 
\begin{equation}
    \vec{v}_{1_N} = \vec{v}_1 -  {v}_{1_L} \dfrac{\vec{r}}{r}; \mbox{ }\mbox{ }\mbox{ } \vec{v}_{2_N} = \vec{v}_2 -  {v}_{2_L} \dfrac{\vec{r}}{r}.
\end{equation} 
The longitudinal $\Delta v_L (r)$ and transverse $\Delta v_N (r)$ velocity differences are then given by
\begin{equation}
    \Delta v_L  (r)=\left|{v}_{1_L} - {v}_{2_L}\right|,
\end{equation}
\begin{equation}
    \Delta v_N (r) = \left| \left(\vec{v}_{1_N}-\vec{v}_{2_N}\right)\cdot\dfrac{\vec{v}_{1_N}}{||\vec{v}_{1_N}||}\right|,
\end{equation}
where $|\mbox{ } |$ represents the absolute value. Each calculated $\Delta v_L (r)$ and $\Delta v_N (r)$ is classified according to its separation distance $r$ and grouped into logarithmically spaced bins. Additionally, multiple snapshots of the velocity field were considered to improve statistical convergence. We have also checked that the results are independent of the number of bins taken.

\begin{figure}[h!]
     \centering
     \includegraphics[width=0.5\textwidth]{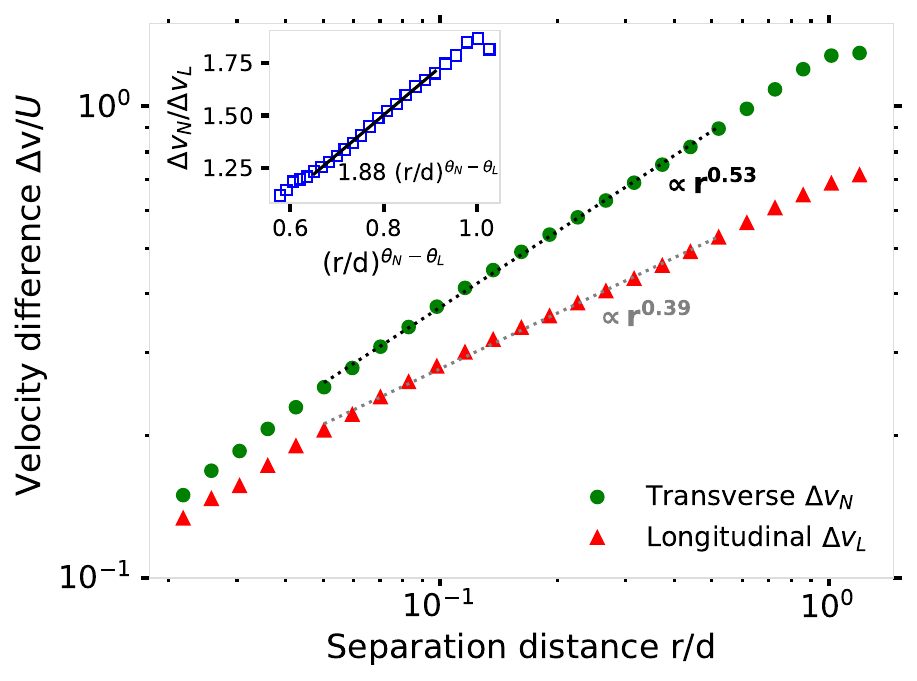}
\caption{Transverse $\Delta v_N$ (green) and longitudinal $\Delta v_L$ (red) velocity difference (normalised by $U$) as a function of separation distance $r$ (normalised by $d$) for $\Ra=3.0 \times 10^4$ ($\Re\sim101$) and $\E=0.73 \times 10^{-3}$. The dotted lines on the curves represent the best fit scaling in the inertial range, $\Delta v_N \propto r^{0.53}$ (black) and $\Delta v_L \propto r^{0.39}$ (grey). The inset shows ${\Delta v_N}{/\Delta v_L}$ as a function of ${(r/d)^{\theta_N-\theta_L}}$ wherein the black solid line represents the best-fit, ${\Delta v_N}/{\Delta v_L} = 1.88 \, {(r/d)^{\theta_N-\theta_L}}$.}
\label{fig: CorrelationExp}
\end{figure} 

Using the outlined procedure, we show in Fig.~\ref{fig: CorrelationExp} the transverse $\Delta v_N$ (green) and longitudinal $\Delta v_L$ (red) velocity differences normalised by the rms velocity $U$ as a function of the separation distance $r$ (normalised by $d$) for $\Ra$ = $3.0 \times 10^4$ and $\E$ = $0.73 \times 10^{-3}$. The result corresponds to the constant $\Ra^*$ case having the highest Reynolds number among all our simulations ($\Re\sim 101$). The dotted lines on the curves represent the best-fit scaling in the inertial range $ r \in [0.05d,0.50d]$, from which we determine the exponents $\theta_L \approx 0.39 \pm0.03$ and $\theta_N \approx 0.53\pm0.02$. The difference in the exponents indicates a deviation from the standard Kolmogorov scaling due to anisotropy in the turbulent velocity field \cite{Romano_Antonia_2001}. This reflects the influence of rapidly rotating convective turbulence and implies a growth rate different from the isotropic case studied in the literature \cite{schekochihin2004simulations}. It is for this reason that we normalise $\gamma$ by $U/d$ as a consistent reference timescale across all simulations.

Finally, the parameter $\zeta$ is obtained by fitting the relation
\begin{equation}
\frac{\Delta v_N}{\Delta v_L}
= \zeta \left( \frac{r}{d} \right)^{\theta_N - \theta_L},
\quad    0.05d<r <0.50d.
\end{equation}
The solid black line in the inset of Fig. \ref{fig: CorrelationExp} shows the corresponding best-fit, with $\zeta\approx 1.88 \pm 0.03$. The values of $\theta_L$, $\theta_N$, and $\zeta$ obtained will now be used in equations \eqref{tl} and \eqref{tn} to model the two-point velocity correlation functions.

\subsection{Growth rate dependence on $\Pm$}\label{sec: KK3}
 
With the small-scale hydrodynamic turbulence in the PNS quantified by the three parameters $\theta_L$, $\theta_N$, and $\zeta$, we now solve the Kazantsev model, equation \eqref{kaz}, to get the growth rate as a function of $\Pm$. A central second-order finite difference scheme is used for discretisation in the three domains. The inner domain is given by $r\in [0, l_c]$, the intermediate region is defined as $ r\in [l_c, d]$ while the outer domain is given by $ r\in [d, r_o]$. We take $r_o = 10 d$ to represent the boundary at infinity. We have verified that the results remain unchanged when $r_o$ is increased further. Additionally, the number of discretisation points is increased as we increase $\Pm$. Solutions are considered to be independent of resolution when the results vary by less than $5\%$ as the resolution is doubled. Homogeneous Dirichlet boundary conditions are implemented, that is, $\psi(r=0)= \psi(r=r_o) = 0$, while the continuity of $\psi$ and $\psi'$ is imposed at $r=l_c$ and $r=d$. The obtained eigenvalue $\Gamma$ characterises the exponential growth of the magnetic field amplitude, therefore, to enable a direct comparison with DNS, we use the magnetic energy growth rate $\gamma = 2\Gamma$.

\begin{figure}[h!]
\centering
\includegraphics[width=0.5\textwidth]{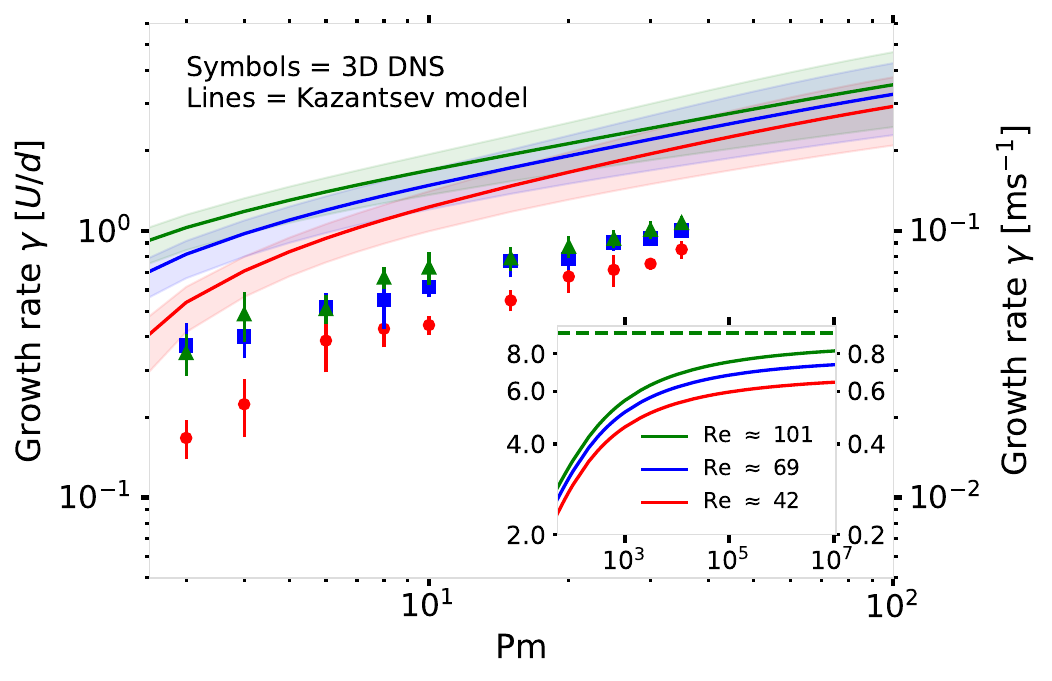}\hfill
\caption{Magnetic energy growth rate $\gamma$ (in units of $U/d$) as a function of $\Pm$. The y-axis on the right is in units of $\si{\per\milli\second}$. The symbols represent the constant $\Ra^*$ set of simulations (same as Fig.~\ref{fig: GammaRoConstant}). The lines are the growth rates from the Kazantsev model for increasing $\Re$, with $\theta_L\approx 0.39$, $\theta_N\approx 0.53$, and $\zeta\approx 1.88$. The shaded regions represent the variation in $\gamma$ due to varying $\Re$ and the fitting uncertainty in $\theta_L$, $\theta_N$, and $\zeta$. The inset is a continuation of the three lines representing $\gamma$ at higher $\Pm$. The dashed line corresponds to the analytical WKB value of $\gamma$ computed at $\Re = 101$ and $\Pm = 10^7$ -- see equation~\eqref{anagamma}.}
\label{Re}
\end{figure} 

Figure \ref{Re} shows the magnetic energy growth rate $\gamma$ as a function of $\Pm$. The symbols represent the constant $\Ra^*$ set of the DNS (same as Fig.~\ref{fig: GammaRoConstant}). The lines correspond to $\gamma$ obtained upon solving the Kazantsev model. All the curves have the same parameters $\theta_L$, $\theta_N$, and $\zeta$, determined by fitting the highest $\Re$ simulation ($\Re\sim101$, see Fig. \ref{fig: CorrelationExp}), where a broad inertial range exists. In the model, increasing $\Re$ widens the inertial range and reduces the viscous scale. The variation in $\gamma$ due to varying $\Re$ and the fitting error in $\theta_L$, $\theta_N$, and $\zeta$ is shown by the shaded regions. The inset is a continuation of the three curves representing $\gamma$ at higher $\Pm$. We find that the Kazantsev model captures the qualitative dependence of $\gamma$ on Pm and Re, but systematically overestimates the growth rate compared to DNS by a factor of $\sim 2.5$. This behaviour is in agreement with earlier studies, which likewise report larger values of $\gamma$ predicted by the Kazantsev model \cite{mason2011magnetic,federrath2016magnetic,schekochihin2004simulations}. The discrepancy can be traced to the idealised assumption of a delta-correlated velocity. As shown by \cite{schekochihin2001finite,bhat2014fluctuation}, finite time correlations inherent to realistic turbulent flows reduce the efficiency of random stretching relative to the ideal model. 

Keeping in mind the difference between the DNS results and the Kazantsev model predictions, we extrapolate our results to higher $\Pm$ values within the framework of the ideal Kazantsev model as shown in the inset of Fig.~\ref{Re}. We observe that $\gamma$ begins to saturate at higher $\Pm$, following a similar trend across all $\Re$. This indicates that the growth rate approaches an asymptote that becomes independent of $\Pm$. However, the value of this asymptote shows a dependence on $\Re$, which can be understood from the scaling of $\gamma$ in the region of strongest shear. Following Schober \cite{schober2012magnetic} and applying the WKB method (details mentioned in Appendix \ref{App:LargePm}), we obtain $\gamma$ in the large $\Pm$ limit as, 
\begin{equation}\label{anagamma}
\gamma = 2\left( \frac{2}{3} \zeta  \Re^{\frac{1-\theta_N}{1+\theta_N}} - \frac{\pi^2}{12 \ln{\Pm}} \Re^{\frac{1-\theta_L}{1+\theta_L}}\right) \dfrac{U}{d} ,
\end{equation}
where the first term inside the bracket represents the asymptotic value as $\Pm \rightarrow \infty$, while the second term is its first-order correction. This indicates that, in the large $\Pm$ regime, $\gamma$ is governed at leading-order by stretching from viscous-scale eddies, while the correction captures the dependence on the resistive cutoff \cite{Kulsrud1992}, resulting in a nontrivial dependence on both scales \cite{xu2016turbulent, schekochihin2002small, schober2012magnetic, schekochihin2004simulations}. The correction remains small but non-negligible even at large $\Pm$, owing to its $1/\ln(\Pm)$ scaling, implying that $\gamma$ is only weakly sensitive to resistivity in a PNS (see Appendix \ref{App:LargePm} for details). As shown in the inset of Fig.~\ref{Re}, for $\Re \sim 101$, the saturating value of $\gamma$ at $\Pm = 10^7$ differs by $\sim 10\%$ from the corresponding analytical WKB prediction indicated by the horizontal dashed line. This difference is expected to decrease as we move into the large $\Pm$ regime, where higher-order corrections become insignificant and thus, equation \eqref{anagamma} accurately captures the underlying dynamics. For the physical conditions relevant to PNS, that is, large $\Pm \sim 10^{13}$ and high $\Re \sim 10^4$ \cite{lander2021generating}, the WKB method applied to the Kazantsev model therefore predicts a growth rate 
\begin{equation}\label{gamma}
    \gamma \sim 40 (U/d)\sim 4 \left(\frac{U}{{10^{5}}\si{\centi\meter \milli\per\second}}\right) \left(\frac{10^6\si{\centi\meter}}{d}\right) \si{\per\milli\second},
\end{equation}
suggesting that the small-scale dynamo is highly efficient in a PNS, enabling a rapid amplification of the magnetic field. Assuming that the model overpredicts $\gamma$ by a factor of $\sim 2.5$ even at large $\Pm$, we still obtain a very fast growth rate of the order of $\SI{1}{ms^{-1}}$. Hence, we can conclude that the growth phase of the magnetic energy is much shorter than the duration of PNS convection.

It is essential to emphasise that the theoretical model employed here represents a one-dimensional approximation of the three-dimensional DNS setup. Additionally, the effects of rotation of the PNS are not taken into account, and homogeneity in the fields is assumed to be valid. The model, therefore, is only an approximate description of the magnetohydrodynamic turbulence in PNSs. Despite its simplifications, the ideal Kazantsev model captures the qualitative behaviour with $\Pm$ and $\Re$ and helps to predict the growth rate at large $\Pm$ relevant to PNS.

\section{Conclusion} \label{sec: conclusion}
To study the magnetic field amplification during the short convective phase in a PNS, 82 3D DNS of convective dynamo are performed under the anelastic approximation. In this paper, we focus only on the kinematic phase of the dynamo. We find that the growth rate $\gamma$ of the magnetic energy increases with both $\Pm$ and $\Ra$, but smoothly transitions to a weaker scaling at high $\Pm$. Simultaneously, the axisymmetric and dipolar components of the magnetic energy decrease as $\Rm^{-1}$, making the fastest-growing mode strongly non-axisymmetric and multipolar. It is interesting to note that the large-scale dynamo wave distinctly observed for $\Rm \lesssim 10^3$ is completely lost beyond $\Rm \sim 2\times 10^3$ for $\Ro \in [0.07,0.15]$, indicating a gradual transition from a large-scale to a small-scale dynamo. Consequently, the magnetic field is dominated by small-scale structures as $\Rm$ increases, while the kinetic energy scales remain approximately constant, leading to a gradually increasing scale separation between the two. These results are found to be independent of the outer magnetic boundary condition. 

We extend our results to the physically relevant but computationally inaccessible regime of large $\Pm$ and high $\Re$ in PNSs, using the theoretical small-scale Kazantsev dynamo model. A salient feature of our approach is that the PNS turbulence is modelled using parameters constrained by DNS and consistently incorporated into the theoretical framework to account for deviations from isotropy in fast-rotating convective turbulence as well as the effects of compressibility. We find that the model qualitatively captures the dependence of the growth rate on $\Pm$ and $\Re$ observed in DNS, but systematically overestimates it by a factor of 2.5. This overestimate, seen in earlier studies, can be traced to the idealised assumption of a delta-correlated velocity. Keeping in mind the difference, we extrapolate the results to higher $\Pm$ values. We find that $\gamma$ begins to saturate towards an asymptote that depends on $\Re$. To interpret this behaviour, we derive the standard WKB analytical solution to the Kazantsev model in the large $\Pm$ limit, which predicts a growth rate of $\sim \SI{4}{\per\milli\second}$ under PNS conditions. It is important to highlight that this $\gamma$ value is found to exhibit only a weak dependence on the resistivity in a PNS. Assuming that the model over-predicts $\gamma$ by a factor of $\sim 2.5$ even at large $\Pm$, we still obtain a very fast growth rate of the order of $\SI{1}{\per\milli\second}$, implying amplification on timescales much shorter than the duration of PNS convection. Overall, this provides a first quantitative estimate of the magnetic energy amplification in a PNS and suggests that a highly efficient small-scale dynamo can generate magnetar-strength fields during the short convective phase.

Our results can be placed in the broader context of turbulent dynamo studies. In binary neutron star (BNS) merger simulations, it is observed that magnetic fields undergo rapid amplification to magnetar strengths on millisecond timescales, driven by the Kelvin–Helmholtz and magneto-rotational instabilities \cite{Price2006, Giacomazzo2015}. The associated growth rates obtained are of the order of $\sim 1~\mathrm{ms}^{-1}$ \cite{Kiuchi2015, Kiuchi2018}. Despite the different driving mechanisms, localised shear in BNS mergers versus global buoyancy in the PNS, the similar short timescales indicate a common feature of magnetic energy amplification through small-scale dynamo action in neutron star environments. In the context of rotating convective dynamos, parameter-dependent transitions in magnetic field topology are well established \cite{aubert2008, olson2006}. Dipole-dominated and axisymmetric fields are typically found at low $\Ro$, while increasing $\Ro$ and $\Ra$ lead to multipolar and non-axisymmetric fields \cite{ChristensenAubert2006, gastine2012dipolar}. This is consistent with the reduction of dipolar and axisymmetric components seen in our results. However, the predominance of small-scale magnetic energy by the end of the growth phase is inconsistent with the strong surface dipolar field of magnetars. If the inferred large-scale magnetic field of magnetars originates from a convective dynamo, one possible scenario is that it undergoes further amplification during the nonlinear phase \cite{raynaud2020magnetar}. More generally, at high $\Rm$, large-scale field generation is typically enabled by helicity, shear, or nonlinear transfer processes \cite{tobias2021turbulent}. In this context, dynamos driven by magneto-rotational instability in the high-Pm regime are actively being investigated \cite{guilet2022mri, held2022, Kiuchi:2023obe, reboul2021global, reboul2022mri, Reboul-Salze:2025eqt}. A detailed study of the nonlinear saturation of the convective dynamo in PNS will be the subject of future work.

\begin{acknowledgments}
SV acknowledges the support from the Prime Minister’s Research Fellows (PMRF) scheme, Ministry of Education, Government of India. KS and SV would like to thank the computing resources and support provided by PARAM Shakti, the supercomputing facility of IIT Kharagpur, established under the National Supercomputing Mission (NSM) of the Government of India and supported by the Centre for Development of Advanced Computing (CDAC), Pune. JG and SV appreciate the support from the Munich Institute for Astro-, Particle, and BioPhysics (MIAPbP), which is funded by the Deutsche Forschungsgemeinschaft (DFG, German Research Foundation) under Germany´s Excellence Strategy – EXC-2094 – 390783311.
\end{acknowledgments}
\appendix
\section{Large $\Pm$ limit} \label{App:LargePm}
In this section, we derive an analytical solution for the growth rate $\gamma$ in the large $\Pm$ limit using the standard WKB method. Following Schober et al. (2012) \cite{schober2012magnetic}, we can rewrite the Kazantsev equation \eqref{kaz} as
\begin{equation}
        \psi''(r) + p(r) \psi(r) = 0,
    \label{rekaz}
\end{equation}
where the modified potential is given by
\begin{equation}
    p(r) = -\dfrac{V(r)+\Gamma}{\kappa_{\text{d}}(r) }.
    \label{pr}
\end{equation}
At sufficiently large $\Pm$, only the dissipative scales are important. Hence, introducing a new dimensionless variable,
\begin{equation}
    z = \left( \dfrac{U\Re^{\frac{1-\theta_L}{1+\theta_L}}}{3 D \eta} \right)^{1/2} r, 
\end{equation}
and using equations \eqref{tl} and \eqref{tn} in the dissipative regime in the equation \eqref{pr}, $p(z)$ can be written as,
\begin{equation}
    p(z)=\dfrac{A_0z^4 - B_0z^2 - 2}{z^2 (1+z^2)^2},
\end{equation}
with,
\begin{gather}
    A_0 = 2\zeta\Re^{\frac{1-\theta_L}{1+\theta_L} - \frac{1-\theta_N}{1+\theta_N}}-\bar{\Gamma }\dfrac{\Re\Pm}{\tilde{R}^2}, \\
    B_0 = 3- 2\zeta\Re^{\frac{1-\theta_L}{1+\theta_L} - \frac{1-\theta_N}{1+\theta_N}} +\bar{\Gamma }\dfrac{\Re\Pm}{\tilde{R}^2},
\end{gather}
where, $\tilde{R}^2 = \dfrac{U D}{3 \eta}\Re^{\frac{1-\theta_L}{1+\theta_L}}, \mbox{ and }\bar{\Gamma} = \dfrac{D}{U} \Gamma$ is the normalised growth rate. In the limit of large $\Pm$, we have,
\begin{equation} \label{pzPm}
    p(z) = \dfrac{A_0z^2 - B_0}{z^4}.
\end{equation}
The real, positive root of this equation, $z_1 = \sqrt{\dfrac{B_0}{A_0}}$, is one turning point. The other turning point is the cutoff scale of turbulence, $z_2 = \dfrac{\tilde{R}}{D}r|_{\ell_c} = \sqrt{\dfrac{Pm}{3}}$. This gives the eigenvalue condition in the WKB approximation as,
\begin{equation}\label{int}
    \int_{z_1}^{z_2} \sqrt{\dfrac{A_0z^2 - B_0}{z^4}}dz = \dfrac{\pi}{2}.
\end{equation}
We can solve the integral in \eqref{int} analytically, and for $z_2 \gg 1$, which is the large $\Pm$ limit, a zero-order iterative solution for the normalised growth rate gives,
\begin{equation} \label{eqn: ana gamma large pm}
\bar{\gamma}=2\bar{\Gamma} = 2\left(\dfrac{2}{3}\zeta \Re^{\frac{1-\theta_N}{1+\theta_N}} - \dfrac{\pi^2}{12 \ln{\Pm}} \Re^{\frac{1-\theta_L}{1+\theta_L}}\right).
\end{equation}
This is the same as mentioned in equation \eqref{anagamma}. Its asymptotic value as $\Pm \rightarrow \infty$, is given by,
\begin{equation}\label{eqn: ana gamma infty pm}
    \bar{\gamma}=2\bar{\Gamma} = \dfrac{4}{3}\zeta \Re^{\frac{1-\theta_N}{1+\theta_N}}.
\end{equation}

\begin{figure}[h!]
     \centering
\includegraphics[width=0.5\textwidth]{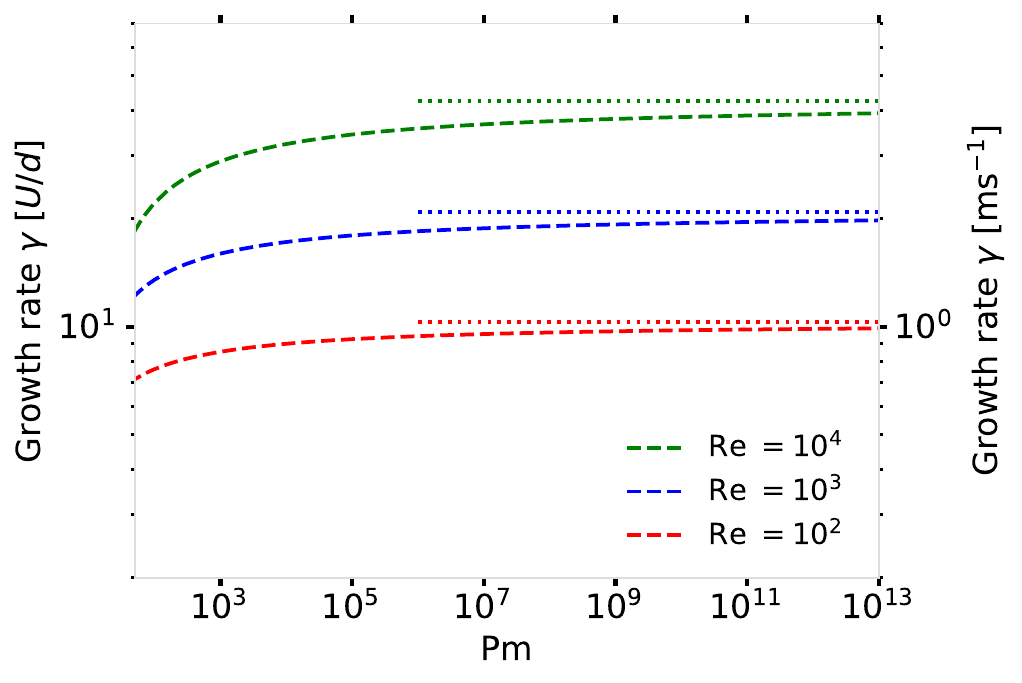}
\caption{Analytical WKB magnetic energy growth rate $\gamma$ in the large $\Pm$ limit (equation \ref{eqn: ana gamma large pm}) as a function of $\Pm$ for different $\Re$. The y-axis on the left is in units of $U/d$ while on the right it is in $\si{\per\milli\second}$. The dotted lines represent the corresponding asymptotic $\gamma$ in the limit $\Pm \rightarrow \infty$ (equation \ref{eqn: ana gamma infty pm}).}
\label{fig: ana gamma infty pm}
\end{figure}

Figure \ref{fig: ana gamma infty pm} shows the WKB magnetic energy growth rate $\gamma$ in the large $\Pm$ limit as a function of $\Pm$, with the dotted lines representing the asymptotic value as $\Pm \rightarrow \infty$. As the first-order correction to the large $\Pm$ limit $\gamma$ scales logarithmically with $\Pm$, it approaches its asymptotic value very slowly. This behaviour is observed across all $\Re$, as shown in the Fig. \ref{fig: ana gamma infty pm}. The green dashed line shows the WKB $\gamma$ prediction at large $\Pm$ for the PNS with $\Re=10^4$. We find that even at $\Pm=10^{13}$, the large-$\Pm$ growth rate has not yet converged to its asymptotic value, indicating that the first-order correction, although small, remains non-negligible due to the 1/$\ln({\Pm})$ dependence. Hence, we can infer that the growth rate of the magnetic energy is only weakly dependent on the resistivity in the PNS environment.

\bibliography{refs}
\bibliographystyle{apsrev}
\end{document}